\documentclass[reprint,prb,showpacs]{revtex4-1}
\newcommand{\ovb}[1]{ \overset{\text{\tiny$\bm\leftrightarrow$}}{\boldsymbol{#1}} \mkern2mu\vphantom{#1}}
\usepackage{amsmath}
\usepackage{amssymb}
\usepackage{graphicx}
\usepackage{slashed}
\usepackage{comment}
\usepackage{bm}
\usepackage{ushort}

\begin{document}

\title{Surface - lattice resonances in 2d arrays of
spheres:\\ multipolar interactions and a mode analysis }

\author{Sylvia D. Swiecicki}
\email{sswiecic@physics.utoronto.ca}
\author{J. E. Sipe}
\affiliation{Department of Physics, University of Toronto, Toronto, Ontario, Canada M5S1A7}


\date{\today}

\begin{abstract}
We present a multipolar model of surface - lattice resonances (SLRs) in 2d arrays of spheres including the electric dipole, magnetic dipole, and electric quadrupole moments of the spheres. We identify SLRs of dipolar and multipolar character, show the importance of non-resonant multipoles in their description, and discuss the sensitivity of SLRs to illumination conditions. We link SLRs to an excitation of modes supported by the array, and we propose a simplified model of the mode dispersion relations that explains the sensitivity of SLRs and the band gap in mode dispersion found at low frequencies. Finally we discuss the resonant features associated with a direct coupling to a mode which can occur in addition to the diffractive coupling signalled by SLRs.

\end{abstract}

\maketitle

\section{Introduction}
\label{Introduction}

A collection of metallic nanoparticles under
optical illumination can exhibit a resonance structure very different from the plasmon resonance of an isolated emitter.\cite{Lukyanchuk2010} The difference arises due to radiative interactions between the nanoparticles. When nanoparticles are arranged on a lattice, collective interactions give rise to rapid variations of the radiation intensity at wavelengths corresponding to the onset of a new diffraction order.\cite{DeAbajo2007} The dip in the reflection spectrum observed when a new diffraction order becomes evanescent -- the so called Rayleigh anomaly -- is often accompanied by a narrow surface-lattice resonance\cite{Auguie2008, Kravets2008,Vecchi2009,Giannini2010} (SLR) exhibiting an asymmetric Fano profile.\cite{Lukyanchuk2010} Surface-lattice resonances have been studied in a wide range of systems, including 1d chains\cite{Zou2004,Markel2005,Liu2012} and 2d arrays\cite{Zou2004b,Humphrey2014, Teperik, Evlyukhin2010,Evlyukhin2012} of plasmonic and dielectric nanoparticles, nanoparticle arrays on a substrate,\cite{Auguie2010,Mousavi2012} and arrays involving few nanoantennas per unit cell.\cite{Mousavi2011, Humphrey2016, Liu2014} 

The analytic description of the surface-lattice resonances usually relies on approximating each plasmonic nanoparticle as an electric dipole\cite{Zou2004, Zou2004b, Markel2005, Auguie2008, Humphrey2014, Teperik,Auguie2010} or a few dipoles per unit cell in the case of a composite nanoparticle;\cite{Mousavi2011, Humphrey2016} in a description of dielectric nanoparticles the resonant magnetic dipole is also included.\cite{Liu2012,Evlyukhin2010} Although for many systems the dipole approximation is sufficient, higher multipole moments contribute to the radiation from  nanoparticles of moderate
sizes.\cite{Han2009, Evlyukhin2012, Alaee2015, Yong2014, Alu2009, Burrows2010,Dejarnette2012,Khlebtsov2008} In arrays of nanoparticles
these higher order contributions can lead to SLRs with very high quality factors, as was observed in  experiments on arrays of plasmonic nanorods\cite{Giannini2010} and predicted theoretically for an array of spheres with large quadrupolarizabilities.\cite{Evlyukhin2012,Burrows2010} 

The  electrodynamics responsible for the multipolar SLRs is still not well characterized, even in systems as simple as an array of spheres. Radiation from an array of spheres can be found essentially exactly using numerical methods, such as the layer-multipole scattering method,\cite{Stefanou1998,Stefanou2000} but these do not give direct physical insight into the nature of the couplings between the nanoparticles. Such an insight can be gained from analytical models that identify the multipolar structure of resonance spectrum. However, previous analytical treatments\cite{Evlyukhin2012} of SLRs associated with higher order multipoles have been limited in their scope, considering only normally incident light, and only those multipole moments of each sphere that contribute significantly to radiation from an isolated sphere. Yet since the radiation from an array of spheres is different from that of an isolated sphere, with the different multipoles now coupled by radiation interactions that can depend strongly on illumination
conditions, multipoles that have a negligible effect on the scattering of an isolated sphere can become important in the response of an array. Thus a more in-depth analysis of SLRs is needed.

In this paper we present an analytic model of the surface-lattice resonances in 2d arrays of gold spheres. We take into account the electric dipole, magnetic dipole, and electric quadrupole moments, both resonant and non-resonant. We analyze the multipolar structure of the radiation spectrum over a broad range of illumination conditions. We show that strong  interactions between multipoles in the vicinity of Rayleigh anomalies result in a significant contribution from moments that would be negligible if spheres were non-interacting. These non-resonant multipoles are especially important in a
description of SLRs associated with higher multipole moments, for which the contribution from a non-resonant magnetic dipole -- and, depending on the angle of incidence, other multipoles as well -- strongly affects the position and profile of the SLR, as well as its very existence.

Surface-lattice resonances arise due to diffractive coupling of incident light to modes supported by the array;\cite{Carron1986, Ohtaka1982, Bendana2009, Campione2014, Voronov}  we refer to the modes for which this diffractive coupling occurs as surface-lattice modes (SLMs). However, the coupling mechanism and a link between properties of SLRs and those of SLMs has not been discussed in detail. This is because the analytic descriptions of SLRs have often only considered light at normal incidence,\cite{Markel2005, Evlyukhin2010, Evlyukhin2012, Teperik} and so even though the dispersion relations of modes supported by arrays of spheres has been studied extensively,\cite{Alu2009,Savelev2014, Fructos2011, Campione2016, Zhen2008, Shore2012} they have not been discussed in connection with SLRs. Here we discuss the connection between the modes and the resonance spectrum in detail. We identify SLMs of dipolar and multipolar character. We discuss the diffractive coupling to those modes that is signalled by SLRs, as well as a direct coupling that is also possible, with a careful consideration of energy balance in the system. SLMs are lossy due both to absorption in the spheres, and to the very radiative coupling that allows access to them by incident light. With the use of a new representation of periodic Green function dyadics\cite{paper2} that allows us to identify exactly
the part of the coupling that leads to the radiative loss, we can study what might be called ``ideal" SLMs by neglecting both the absorption and the radiative loss. The resulting ideal dispersion relations give us direct insight into under what illumination conditions the SLRs appear, in the same way that the ``ideal" dispersion of a surface plasmon at an air/metal interface, even though calculated with the neglect of absorption loss, can give insight into the response of the system under irradiation from
a cladding prism separated from the interface by an air gap.\cite{Barnes2003,Foley2015} Having
linked SLRs to ideal SLMs, we introduce a simplified model of the ideal SLM dispersion relations that explains their novel features, and we then use it to explain the sensitivity of SLRs to coupling and illumination conditions as well as the frequency cut-off below which SLRs are not found; these properties of SLRs are not easily understood without an appreciation of the SLM dispersion relations.

The manuscript is organized as follows. In section \ref{model} we introduce the multipolar model. In section \ref{couplings} we present  resonance spectrum for illumination along lattice symmetry direction. In sec. \ref{normal modes} we find the dispersion relations of the SLMs. In section \ref{simple model} we propose a simplified model of the dispersion relations  and we link it to the properties of SLRs in \mbox{section \ref{discussion}}. In section \ref{other} we analyse the sensitivity of SLRs to the direction and polarization of light. In \mbox{section \ref{substrate}} we discuss the direct coupling to SLMs. We conclude in \mbox{section \ref{conclusions}.}

\section{Multipolar model}
\label{model}

\subsection{Isolated sphere}
\begin{figure}[htb]
\includegraphics[scale=0.3]{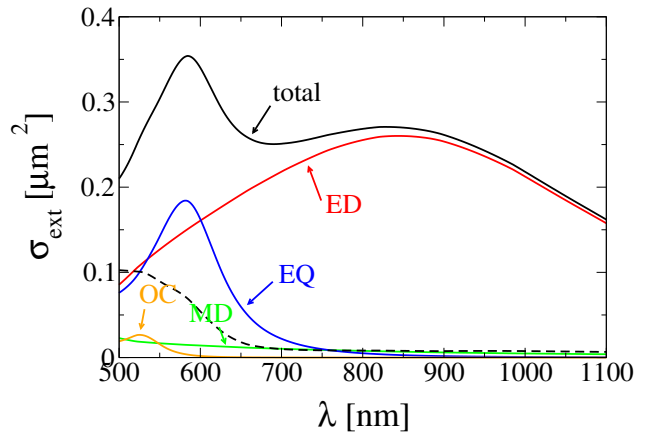}
\caption{Extinction cross section of an isolated sphere and contributions from multipoles: ED - electric dipole, MD - magnetic dipole, EQ - electric quadrupole, OC - electric octupole. Dotted line shows the total absorption cross section.}
\label{isolated}
\end{figure}

As an example of typical nanoparticles we consider gold spheres of radius $r =100nm$. Adopting the optical constants of
Johnson and Christy,\cite{JC}
and assuming the nanoparticles are embedded in a medium with index of
refraction $n=1.45$, we find from Mie theory\cite{BW} that the extinction of an isolated sphere illuminated by a plane wave is dominated by scattering at wavelengths above 600nm (see Fig.~\ref{isolated}).
The main contribution to extinction comes from the electric dipole and the electric quadrupole moment, while the magnetic dipole moment gives only a small non-resonant contribution; see Fig.~\ref{isolated}. Nevertheless, the magnetic dipole moment formally enters the multipole expansion at the same level as electric quadrupole,\cite{Raab} and it is important in describing the collective effects in radiation from a lattice of spheres, as we shall show. We thus  develop a description including
the electric dipole, magnetic dipole, and electric quadrupole moments
of the spheres. We refer to this as the \textit{multipolar model}.

While Mie theory is usually derived in the context of a plane-wave illumination,\citep{Doyle,Markel2005,Alu2009, Evlyukhin2010, Evlyukhin2012} even for an arbitrary incident field the Mie polarizabilities relate the multipole moments to electromagnetic fields and their gradients at the center of the sphere.\cite{Das2015} We consider a response of a sphere to electromagnetic fields of the form
\begin{align}
\boldsymbol{E}^{\text{inc}}(\boldsymbol{r},t) &= \boldsymbol{E}^{\text{inc}}(\boldsymbol{r})e^{-i\omega t}+c.c, \label{Ew}\\
\boldsymbol{B}^{\text{inc}}(\boldsymbol{r},t) &= \boldsymbol{B}^{\text{inc}}(\boldsymbol{r})e^{-i\omega t}+c.c,\label{Bw} 
\end{align}
where $\boldsymbol{E}^{\text{inc}}(\boldsymbol{r})$ in general need not be a plane wave. In what follows we choose to work with multipole moments defined in Cartesian basis. We identify the electric dipole moment $\boldsymbol{p}$, magnetic dipole moment $\boldsymbol{m}$, and electric quadrupole moment $\overset{\text{\tiny$\bm\leftrightarrow$}}{\boldsymbol{q}}$ as  Cartesian versions of the full multipole coefficients, defined in terms of integrals of the charge and current distribution taken with the spherical Bessel functions; these moments describe all the radiation pattern with the polarity up to the electric quadrupole level.\cite{Jackson} Writing $\boldsymbol{E}^{\text{inc}} = \boldsymbol{E}^{\text{inc}}(\boldsymbol{0})$, etc., we have for an isolated sphere at the origin of coordinates,
\begin{align}
\boldsymbol{p} &= \bar{\alpha}^{\text{pE}} \boldsymbol{E}^{\text{inc}},\\
\boldsymbol{m} &= \bar{\alpha}^{\text{mB}} \boldsymbol{B}^{\text{inc}},\\
\ovb{q} &= \bar{\alpha}^{\text{qF}} \ovb{F}^{\text{inc}},
\end{align} 
where the dyadic $F_{ij} = \frac{1}{2}\left( \partial E_j /\partial x_i + \partial E_i /\partial x_j \right)$ is the symmetrized gradient of the electric field, and the Mie polarizabilities are given by\cite{Doyle,Alu2009,Evlyukhin2012}
\begin{align}
\bar{\alpha}^{\text{pE}} &= -\frac{i}{(\tilde{\omega}n)^3} 4\pi\epsilon_0 n^2 B^e_1, \label{pol1} \\ 
\bar{\alpha}^{\text{mB}} &= - \frac{i}{(\tilde{\omega}n)^3} 4\pi\epsilon_0 c^2 B^m_1, \\
\bar{\alpha}^{\text{qF}} &= -\frac{1}{(\tilde{\omega}n)^5} 24\pi\epsilon_0n^2 B^e_2, \label{pol2} 
\end{align}
where $\tilde{\omega}=\omega/c$, and $B^e_l$, $B^m_l$ are the amplitudes of the associated partial waves.\cite{BW} The polarizabilities as defined in (\ref{pol1}-\ref{pol2}) involve contributions that give rise to the radiative damping. To see the energy balance in the system more explicitly we find it more convenient to work with proper polarizabilities, the inverses of which are given by
\begin{align}
\left( \alpha^{\text{pE}} \right)^{-1} &= \left(\bar{\alpha}^{\text{pE}}\right)^{-1} +\frac{i(\tilde{\omega} n)^3}{6\pi\epsilon_0 n^2}, \label{proper1}\\
\left( \alpha^{\text{mB}} \right)^{-1} &= \left(\bar{\alpha}^{\text{mB}}\right)^{-1} +\frac{i(\tilde{\omega} n)^3}{6\pi\epsilon_0 c^2}, \\
\left( \alpha^{\text{qE}} \right)^{-1} &= \left(\bar{\alpha}^{\text{qE}}\right)^{-1} +\frac{i(\tilde{\omega} n)^5}{20\pi\epsilon_0 n^2}, \label{proper2}
\end{align}
where the terms $i(\tilde{\omega} n)^3\boldsymbol{p}/(6\pi\epsilon_0 n^2)$, $i(\tilde{\omega} n)^3\boldsymbol{m}/(6\pi\epsilon_0 c^2)$, and $i(\tilde{\omega} n)^5 \ovb{q}/(20\pi\epsilon_0 n^2)$ are the radiation reaction fields associated with each multipole.\cite{paper2} Identifying the sum of each incident field and the associated radiation reaction field,
\begin{align}
\boldsymbol{E'} &= \boldsymbol{E}^{\text{inc}} + \frac{i(\tilde{\omega}n)^3}{6\pi \epsilon_0 n^2} \boldsymbol{p},\label{rr}\\ 
\boldsymbol{B'} &= \boldsymbol{B}^{\text{inc}} + i\frac{(\tilde{\omega}n)^3}{6\pi \epsilon_0 c^2} \boldsymbol{m},\\
\ovb{F}' &= \ovb{F}^{\text{inc}} + i\frac{(\tilde{\omega}n)^5}{20\pi \epsilon_0 n^2} \ovb{q},
\end{align}
the response of an isolated sphere to fields at frequency $\omega$ is given in terms of the proper polarizabilities by
\begin{align}
\boldsymbol{p} &= \alpha^{\text{pE}} \boldsymbol{E'}, \\
\boldsymbol{m} &= \alpha^{\text{mB}} \boldsymbol{B'}, \\ 
\ovb{q} & = \alpha^{\text{qF}} \ovb{F}'.
\end{align}

\subsection{2D triangular lattice}

\begin{figure}[htb]
\includegraphics[scale=0.35]{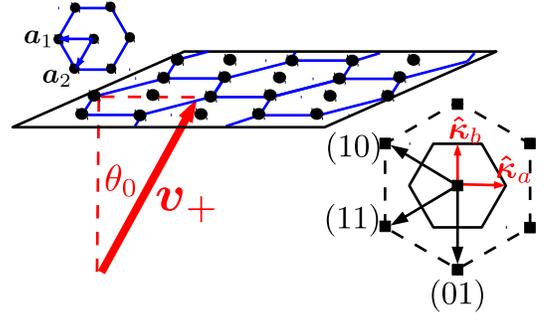}
\caption{Illumination conditions and reciprocal lattice. The incident wave vector is $\boldsymbol{v}_+$, the angle of incidence is $\theta_0$, and the lattice vectors are $\boldsymbol{a}_1$, $\boldsymbol{a}_2$. The  reciprocal lattice, indicated by squares, is spanned by the basis vectors $(10)$ and $(01)$; we also indicate the reciprocal vector (11) explicitly. The black continuous and dashed lines in the diagram of the reciprocal lattice are the boundaries of the first and the second Brillouin zone respectively.}
\label{symmetry}
\end{figure}

We now consider a triangular array of these spheres as sketched in Fig.~\ref{symmetry}, with the basis lattice vectors of length $\left|\boldsymbol{a}_1\right| = \left|\boldsymbol{a}_2\right| = 475nm$. 
The centers of the spheres are taken to lie in the $z = 0$ plane; in this plane we label the coordinates by $\boldsymbol{R} = (x, y)$, and we write the coordinates of the centers of the spheres as $\boldsymbol{R}_{\boldsymbol{n}} = (x_{\boldsymbol{n}} , y_{\boldsymbol{n}} )$ where $\boldsymbol{n}=(n_1,n_2)$ is a vector of integers indexing the lattice sites along the basis vectors, $\boldsymbol{R}_{\boldsymbol{n}} = n_1 \boldsymbol{a}_1 + n_2 \boldsymbol{a}_2$. We assume the spheres are illuminated
by a plane electromagnetic wave with wave vector $\boldsymbol{v}_+ = \boldsymbol{\kappa}_0 + w_0 \boldsymbol{\hat{z}}$, $\boldsymbol{v}_+\cdot \boldsymbol{v}_+ = (\tilde{\omega}n)^2$, 
\begin{align}
\boldsymbol{E}^{\text{inc}}(\boldsymbol{r}) &= \boldsymbol{E}^{\text{inc}} e^{i\boldsymbol{\kappa}_0\cdot \boldsymbol{R}} e^{iw_0z}, \label{Einc} \\
\boldsymbol{B}^{\text{inc}}(\boldsymbol{r}) &= \boldsymbol{B}^{\text{inc}} e^{i\boldsymbol{\kappa}_0\cdot \boldsymbol{R}} e^{iw_0z},\label{Binc}
\end{align}
where $\boldsymbol{\kappa}_0$ is the in-plane component of the incident wave vector, $|\boldsymbol{\kappa}_0| = \tilde{\omega}n\text{sin}\theta_0$,  $\theta_0$ is the angle of incidence, and $w_0 = \sqrt{ (\tilde{\omega}n)^2 - \kappa_0^2 }>0$. 

For this excitation, the induced multipole moments at the lattice site 
$\boldsymbol{R}_{\boldsymbol{n}}$,  $(\boldsymbol{p}_{\boldsymbol{n}},\boldsymbol{m}_{\boldsymbol{n}},\ovb{q}_{\boldsymbol{n}})$, are related to those at the origin, $(\boldsymbol{p},\boldsymbol{m},\ovb{q})$, by
\begin{align}
\boldsymbol{p_n} &= e^{i\boldsymbol{\kappa}_0\cdot \boldsymbol{R_n}} \boldsymbol{p},\label{rel1} \\
\boldsymbol{m_n} &= e^{i\boldsymbol{\kappa}_0\cdot \boldsymbol{R_n}} \boldsymbol{m},\\
\ovb{q}_{\boldsymbol{n}} &= e^{i\boldsymbol{\kappa}_0\cdot \boldsymbol{R_n}} \ovb{q},\label{rel2}
\end{align}
and so the scattering problem can be formulated in terms of $\boldsymbol{p}, \boldsymbol{m}, \ovb{q}$ only. We have
\begin{align}
\boldsymbol{p} &= \alpha^{\text{pE}} \boldsymbol{E}^{\text{tot}},\label{eq1}\\
\boldsymbol{m} &= \alpha^{\text{mB}} \boldsymbol{B}^{\text{tot}},\\
\ovb{q} &= \alpha^{\text{qF}} \ovb{F}^{\text{tot}},\label{eq2}
\end{align}
where $\boldsymbol{E}^{\text{tot}}, \boldsymbol{B}^{\text{tot}}$ and $\ovb{F}^{\text{tot}}$ are sums of the incident fields, the radiation reaction fields of the sphere at the origin, and the fields radiated from all the other multipoles in an array,
\begin{align}
\boldsymbol{E}^{\text{tot}} &= \boldsymbol{E}^{\text{inc}} + \ovb{\mathcal{G}}^{\text{Ep}} \cdot \boldsymbol{p} +\ovb{\mathcal{G}}^{\text{Em}}\cdot \boldsymbol{m} + \ovb{\mathcal{G}}^{\text{Eq}}:\ovb{q},\label{Etot}\\
\boldsymbol{B}^{\text{tot}} &= \boldsymbol{B}^{\text{inc}} +\ovb{\mathcal{G}}^{\text{Bp}}\cdot \boldsymbol{p} +\ovb{\mathcal{G}}^{\text{Bm}}\cdot\boldsymbol{m} + \ovb{\mathcal{G}}^{\text{Bq}}:\ovb{q},\\
\ovb{F}^{\text{tot}} &= \ovb{F}^{\text{inc}} + \ovb{\mathcal{G}}^{\text{Fp}}\cdot \boldsymbol{p} + \ovb{\mathcal{G}}^{\text{Fm}}\cdot \boldsymbol{m} +\ovb{\mathcal{G}}^{\text{Fq}}:\ovb{q},\label{Ftot}
\end{align}
and we refer to the $\ovb{\mathcal{G}}$'s as the ``periodic
Green functions" of the array.\cite{paper2} For example,
\begin{align}
\ovb{\mathcal{G}}^{\text{Ep}}\cdot \boldsymbol{p} &= \lim_{z\rightarrow 0} \sum_{\boldsymbol{n}\neq (0,0)} e^{i\boldsymbol{\kappa}_0\cdot \boldsymbol{R_n}} \ovb{g}^{\text{Ep}}(-\boldsymbol{R_n}+z\boldsymbol{\hat{z}})\cdot \boldsymbol{p}\nonumber\\
&+\frac{i(\tilde{\omega}n)^3}{6\pi \epsilon_0 n^2} \boldsymbol{p}
,\label{G1}
\end{align}
specifies the electric field at the origin from the electric dipoles at all the other lattice sites, together with the radiation reaction field from the dipole at the origin; the dyadic $\ovb{g}^{\text{Ep}}(\boldsymbol{r})$ in the definition (\ref{G1}) is the usual free-space dyadic, where $\ovb{g}^{\text{Ep}}(\boldsymbol{r})\cdot \boldsymbol{p}$ gives the electric field at $\boldsymbol{r}$ due to an electric dipole $\boldsymbol{p}$ located at the origin. The other terms, $\ovb{\mathcal{G}}^{\text{Bm}}\cdot \boldsymbol{m}$ and $\ovb{\mathcal{G}}^{\text{Fq}}:\ovb{q}$ respectively, give the magnetic field at the origin from the magnetic dipoles at all the other lattice sites together with the radiation reaction field from magnetic dipole at the origin, and the symmetrized field gradient at the origin due to the electric quadrupoles at all the other lattice sites together with the radiation reaction from quadrupole at the origin. All the remaining periodic Green functions are defined as sums of fields at origin from corresponding multipoles at other lattice sites, but do not involve radiation reaction terms; all the periodic Green functions are implicit functions of \mbox{$\boldsymbol{\kappa}_0$ and $\omega$.}

We evaluate the periodic Green function dyadics that enter eqs. (\ref{Etot}-\ref{Ftot}) using a method developed earlier.\cite{paper2} The method gives exact expressions for the radiative contributions to the periodic Green functions, which are responsible for establishing the energy balance of the system, and expressions for the non-radiative contributions in a form of rapidly converging sums that can be numerically evaluated. Each periodic Green function involves Fourier contributions associated with each of the reciprocal lattice vectors
$\boldsymbol{K}$, with non-analyticities as a function of $\boldsymbol{\kappa}_0$ associated with the appearance of each new diffracted order; these occur when $\tilde{\omega}n = \left|\boldsymbol{\kappa}_0 + \boldsymbol{K}\right|$.
Singular behavior in the dyadics as functions of $\boldsymbol{\kappa}_0$ arises because of their dependence, for all $\boldsymbol{K}$, on $\left[\tilde{\omega}^2 n^2 - \left(\boldsymbol{\kappa}_0 + \boldsymbol{K}\right)^2 \right]^{-1/2}$. Terms of this nature arise from the confinement of the radiating multipole moments to a plane, and they play a dominant role in the formation of SLRs and the modes associated with them.\cite{Markel2005, Mousavi2011}

Equations (\ref{eq1}-\ref{eq2}) take a particularly simple form for an incident plane wave with wave vector component in the plane of the lattice, $\boldsymbol{\kappa}_0$, lying along one of the two symmetry directions of the lattice. These two excitation scenarios correspond to $\boldsymbol{\kappa}_0$ in the direction of  $\boldsymbol{\hat{\kappa}}_a$ or in the direction of $\boldsymbol{\hat{\kappa}}_b$; see Fig.~\ref{symmetry}. Note that for the first direction the wave vector $\boldsymbol{\kappa}_0$ is aligned with
one of the lattice vectors, while in the second it is aligned with one of the reciprocal lattice vectors. For $\boldsymbol{\kappa}_0$ along either of these directions the self-consistent equations (\ref{eq1}-\ref{eq2}) decouple into two independent sets of equations. We consider the specific form that these equations take for s- and p-polarized incident light. We define the s- and p-polarization amplitudes in a usual way,
\begin{align}
\boldsymbol{E}^{\text{inc}} = E^{\text{inc}}_s\boldsymbol{\hat{s}} +E^{\text{inc}}_p \boldsymbol{\hat{p}}_{+},\label{sp}
\end{align}
where for a plane wave with an in-plane wave vector component $\boldsymbol{\kappa}$ we define the polarization vectors in general as
\begin{align}
\boldsymbol{\hat{s}} &= \boldsymbol{\hat{\kappa}}\times \boldsymbol{\hat{z}},\label{svec}\\
\boldsymbol{\hat{p}}_{\pm} &= (\tilde{\omega}n)^{-1} \left[ \kappa\boldsymbol{\hat{z}} \mp w\boldsymbol{\hat{\kappa}} \right],\label{pvec} 
\end{align} 
with $w = \sqrt{(\tilde{\omega}n)^2-\boldsymbol{\kappa}^2}$, $\text{Im}(w)>0$, but for use in (\ref{sp}) and the following equations in this paragraph we take $\boldsymbol{\kappa} = \boldsymbol{\kappa}_0$ in (\ref{svec},\ref{pvec}). For s-polarized incident light and $\boldsymbol{\kappa}_0$ along either the direction $\boldsymbol{\hat{\kappa}}_a$ or $\boldsymbol{\hat{\kappa}}_b$, the equations (\ref{eq1}-\ref{eq2}) reduce to the form
\begin{align}
\mathbb{S} \left(\begin{array}{c} 
p_s \\
m_z \\
q_{\kappa s}
\end{array} \right) = \left(\begin{array}{c} 
E^{\text{inc}}_s \\
B^{\text{inc}}_z \\
F^{\text{inc}}_{\kappa s} 
\end{array}\right),\label{S1}
\end{align}
and 
\begin{align}
\mathbb{S}' \left(\begin{array}{c} 
m_\kappa \\
q_{sz}
\end{array} \right) = \left(\begin{array}{c} 
B^{\text{inc}}_{\kappa} \\
F^{\text{inc}}_{sz} 
\end{array}\right),\label{S2}
\end{align}
where expressions for the matrices $\mathbb{S}$ and $\mathbb{S}'$ are given in the Appendix A; their elements take on different values depending on whether $\boldsymbol{\hat{\kappa}}_0 = \boldsymbol{\hat{\kappa}}_a$ or $\boldsymbol{\hat{\kappa}}_0 = \boldsymbol{\hat{\kappa}}_b$. Similarly for p-polarized light the equations reduce to
\begin{align}
\mathbb{P} \left(\begin{array}{c} 
p_{\kappa} \\
q_{\kappa \kappa} \\
q_{ss} 
\end{array} \right) = \left(\begin{array}{c} 
E^{\text{inc}}_{\kappa} \\
F^{\text{inc}}_{\kappa \kappa} \\
0 
\end{array}\right),\label{P1}
\end{align}
and 
\begin{align}
\mathbb{P}' \left(\begin{array}{c} 
p_z \\
m_s \\
q_{\kappa z}
\end{array} \right) = \left(\begin{array}{c} 
E^{\text{inc}}_{z} \\
B^{\text{inc}}_{s} \\
F^{\text{inc}}_{\kappa z} 
\end{array}\right),\label{P2}
\end{align}
where expressions for the matrices $\mathbb{P}$ and $\mathbb{P}'$ are given in the Appendix A; again their elements take on different values depending on whether $\boldsymbol{\hat{\kappa}}_0 = \boldsymbol{\hat{\kappa}}_a$ or $\boldsymbol{\hat{\kappa}}_0 = \boldsymbol{\hat{\kappa}}_b$

Once the multipole moments $\boldsymbol{p},\boldsymbol{m}, \ovb{q}$ are found, we calculate the electromagnetic fields radiated by an array using the Green function formalism for s- and p-polarized light.\cite{Sipe1987,paper2}  Introducing the Fourier transform of a function $\boldsymbol{\mathcal{O}}(\boldsymbol{r})$ in the x,y plane,
\begin{equation}
\boldsymbol{\mathcal{O}}(\boldsymbol{r}) = \int\frac{d\boldsymbol{\kappa}}{(2\pi)^2} \boldsymbol{\mathcal{O}}(\boldsymbol{\kappa},z) e^{i\boldsymbol{\kappa}\cdot \boldsymbol{R}},
\end{equation}
the electric field scattered by the array, $\boldsymbol{E}^{\text{sc}}(\boldsymbol{\kappa},z)$, is given by
\begin{align}
\boldsymbol{E}^{\text{sc}}(\boldsymbol{\kappa},z) = \frac{(2\pi)^2}{A_c} \sum_{\boldsymbol{n}} \delta(\boldsymbol{\kappa}-\boldsymbol{\kappa}_{\boldsymbol{n}}) \boldsymbol{f}^{\mathrm{E}}(\boldsymbol{\kappa},z),\label{EA}
\end{align}   
where $A_c$ is the area of a unit cell, and $\boldsymbol{n}$ here and henceforth indicates a vector of integers indicating directions in reciprocal space; earlier, at the start of this section, we had used it to indicate directions in real space. We indicate the reciprocal lattice vector identified by $\boldsymbol{n}$ as $\boldsymbol{K_n}$, and write $\boldsymbol{\kappa}_{\boldsymbol{n}} = \boldsymbol{\kappa}_0+\boldsymbol{K} _{\boldsymbol{n}}$ for translation of the component of the incident wave vector in the lattice plane, $\boldsymbol{\kappa}_0$, by the reciprocal lattice vector $\boldsymbol{K}_{\boldsymbol{n}}$. The vector $\boldsymbol{f}^{\mathrm{E}}(\boldsymbol{\kappa},z)$ is defined as
\begin{align}
\boldsymbol{f}^{\mathrm{E}}(\boldsymbol{\kappa},z) &= \ovb{g}^{\text{Ep}}(\boldsymbol{\kappa},z)\cdot \boldsymbol{p} + \ovb{g}^{\text{Em}}(\boldsymbol{\kappa},z)\cdot \boldsymbol{m}\nonumber\\
&+\ovb{g}^{\text{Eq}}(\boldsymbol{\kappa},z):\ovb{q},\label{FE}
\end{align}
where $\ovb{g}^{\text{Ep}}(\boldsymbol{\kappa},z)$, $\ovb{g}^{\text{Em}}(\boldsymbol{\kappa},z)$ and $\ovb{g}^{\text{Eq}}(\boldsymbol{\kappa},z)$ are the Fourier transforms in the plane of the array of the usual free-space Green functions dyadics $\ovb{g}^{\text{Ep}}(\boldsymbol{r})$, $\ovb{g}^{\text{Em}}(\boldsymbol{r})$, and $\ovb{g}^{\text{Eq}}(\boldsymbol{r})$.\cite{paper2} The expressions for the reflection and transmission of an array immediately follow from the expressions for the incident electric field (\ref{Einc}) and the field scattered by an array (\ref{EA}),
\begin{align}
R &= \sum_{\boldsymbol{n} \forall  |\boldsymbol{\kappa}_{\boldsymbol{n}}|<\tilde{\omega}n} R_{\boldsymbol{n}}, \\
T &= \sum_{\boldsymbol{n} \forall |\boldsymbol{\kappa}_{\boldsymbol{n}}|<\tilde{\omega}n} T_{\boldsymbol{n}},
\end{align}
where the functions
\begin{align}
R_{\boldsymbol{n}} &= \frac{1}{|\boldsymbol{E}^{\text{inc}}|^2} \frac{w_{\boldsymbol{n}}}{w_0} \left|\frac{1}{A_c} \boldsymbol{f}^{\mathrm{E}}(\boldsymbol{\kappa}_{\boldsymbol{n}},0_-) \right|^2, \label{Rn}\\
T_{\boldsymbol{n}} &= \frac{1}{|\boldsymbol{E}^{\text{inc}}|^2}\frac{w_{\boldsymbol{n}}}{w_0} \left|\frac{1}{A_c} \boldsymbol{f}^{\mathrm{E}}(\boldsymbol{\kappa}_{\boldsymbol{n}},0_+) \right|^2 \nonumber\\
&+\left[ 1+ \frac{2}{A_c}\frac{\text{Re}  \left(\boldsymbol{E}^{\text{inc}}\right)^* \cdot \boldsymbol{f}^{\mathrm{E}}(\boldsymbol{\kappa}_0,0_+)}{|\boldsymbol{E}^{\text{inc}}|^2}  \right]\delta_{\boldsymbol{n},0},  \label{Tn}
\end{align}
describe the contribution to reflection and transmission from a beam diffracted at an angle $\theta_{\boldsymbol{n}} = \text{sin}^{-1} \left[|\boldsymbol{\kappa}_{\boldsymbol{n}}|/(\tilde{\omega}n)\right]$ with respect to the normal, the specular reflection and transmission are the components $R_{\text{spec}} = R_0$ and $T_{\text{spec}} = T_0$ respectively, and we identified $w_{\boldsymbol{n}} = \sqrt{(\tilde{\omega}n)^2-\boldsymbol{\kappa}_{\boldsymbol{n}}^2} $, $\text{Im}w_{\boldsymbol{n}}>0$, as the z-component of a wave vector with the in-plane component $\boldsymbol{\kappa}_{\boldsymbol{n}}$.

\section{Resonance spectrum and the multipolar couplings}
\label{couplings}

The radiation spectrum of an array of emitters depends on the properties of an isolated emitter as well as on the collective radiative interaction between emitters on a lattice. The collective interactions are especially pronounced at wavelengths close to the onset of diffraction, in the vicinity of which narrow and asymmetric surface-lattice resonances (SLRs) are observed.\cite{Auguie2008, Kravets2008,Vecchi2009,Giannini2010} When the optical response of the system is dominated by one multipole moment,\cite{Markel2005,Zou2004,Teperik} or the multipole moments can be treated as independent,\cite{Evlyukhin2010,Evlyukhin2012,Liu2012} the analysis of SLRs can be greatly simplified by introducing an effective polarizability of an array that incorporates both the single particle and the collective interactions. When the coupling between the multipole moments cannot be neglected, the analysis becomes more complicated. 

Here we analyze the SLRs with the multipolar model (\ref{eq1}-\ref{eq2}), including electric dipole, magnetic dipole, and electric quadrupole moments. For different excitation geometries we consider angle scans at fixed frequencies rather than considering the more usual frequency dependence of the resonant spectrum.\cite{Markel2005,Zou2004,Zou2004b,Evlyukhin2010,Evlyukhin2012} This allows us to distinguish more clearly between the single emitter and the collective characteristics of the resonance spectrum; the polarizability of an isolated emitter for a chosen illumination scenario depends only on the wavelength of light, and thus the resonances observed when varying the incidence angle are driven by collective effects that are described by variations in the periodic Green functions.

For an arbitrary direction of $\boldsymbol{\kappa}_0$ the behavior of the specularly reflected and transmitted light, and that of the diffracted light, can be very complicated. To focus on the underlying physics, in this section and the following three we consider s-polarized light incident along the first symmetry direction of the lattice, $\boldsymbol{\kappa}_0\propto \boldsymbol{\hat{\kappa}}_a$. For light incident along this direction multipole moments are given by the decoupled set of equations (\ref{S1}-\ref{S2}). This simplifies the description of the system, but the scenario is still rich enough to exhibit SLRs associated with both the dipole moments and the higher order moments of the spheres. While a decoupled set of equations also holds for light incident along the other symmetry direction of the lattice, $\boldsymbol{\kappa}_0\propto \boldsymbol{\hat{\kappa}}_b$, in that direction the response is more one-dimensional in nature, as we see in detail in section \ref{substrate}; in the direction $\boldsymbol{\kappa}_0\propto \boldsymbol{\hat{\kappa}}_a$ the
two-dimensional nature of the lattice plays a central role in the behavior of the SLRs and the dispersion relations of modes associated with them, as we see in this section and the next. The extension to more general directions of the incident light, and to p-polarized light, is discussed in section \ref{other}.

\begin{figure}[htb]
\includegraphics[scale=0.3]{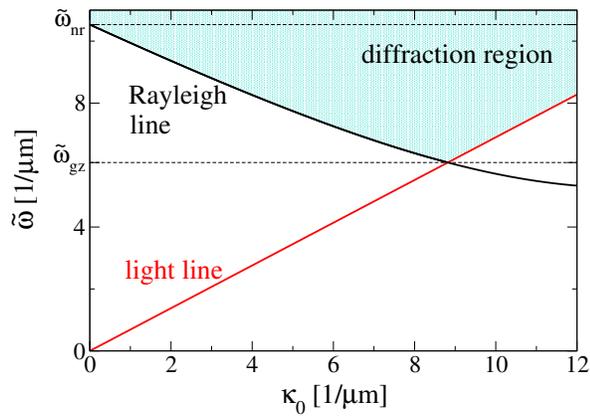}
\caption{The light line (red) and the Rayleigh line (black) for light incident with wave vector component in the plane $\boldsymbol{\kappa}_0$ in the direction $\boldsymbol{\hat{\kappa}}_a$. The shading indicates the incident plane waves, specified by $(\kappa_0, \tilde{\omega})$, for which diffraction is possible. We consider resonance spectra in frequency range restricted by the two dotted lines.}
\label{ldisp}
\end{figure}

First we discuss the range of wavelengths that we will consider. We identify a ``light line" as the condition $\tilde{\omega} = \kappa_0/n$, characterizing light propagating parallel to the plane of the array. This is plotted in Fig.~\ref{ldisp}, where the part of the plane to the left of the light line is accessible to incident light. For $\boldsymbol{\kappa}_0$ in the direction of $\boldsymbol{\hat{\kappa}}_a$, the first of the diffracted orders to appear as the angle of incidence is increased are (10) and (11) (see Fig.~\ref{symmetry}),
occurring at angles $\theta_{(10)}$ and $\theta_{(11)}$ that satisfy the Rayleigh condition with $\boldsymbol{K}=\boldsymbol{K}_{(10)}$ and $\boldsymbol{K}=\boldsymbol{K}_{(11)}$ respectively,
\begin{align}
&|\boldsymbol{\hat{\kappa}}_a \tilde{\omega}n\text{sin}\theta_{\text{(10)}}  + \boldsymbol{K}_{(10)}| = \tilde{\omega}n, \label{tr1}\\
&|\boldsymbol{\hat{\kappa}}_a \tilde{\omega}n\text{sin}\theta_{\text{(11)}}  + \boldsymbol{K}_{(11)}| = \tilde{\omega}n.\label{tr2}
\end{align}
We note that due to symmetry the angles satisfy $\theta_{(10)} = \theta_{(11)} \equiv \theta_R$. The incident angle at which this occurs depends on the wavelength of incident light, and is identified in Fig.~\ref{ldisp} by the ``Rayleigh line", defined as the set of points $(\kappa_0, \tilde{\omega})$ such that $\kappa_0 = \tilde{\omega}n \text{sin}\theta_R$. For fixed wavelength, as the angle of incidence is increased and the Rayleigh line is crossed, the periodic Green functions exhibit singular behaviour. At the Rayleigh line the specular reflection typically, but not always, vanishes. In the standard theory of diffraction from gratings, that kind of behaviour at the onset of diffraction is referred to as a ``Rayleigh anomaly",\cite{RA1,Hessel1965} and we adopt that notation here. For large enough wavelengths, where the Rayleigh line is to the right of the light line, no diffraction is possible; the cut-off for our system is at the wavelength $\lambda_{\text{gz}}\approx 1033nm$ ($\tilde{\omega}_{\text{gz}}\equiv 2\pi/\lambda_{\text{gz}}$), where the diffraction would only arise at grazing incidence, with $\theta_R=90^0$. At $\lambda_{\text{nr}}\approx 596nm$ ($\tilde{\omega}_{\text{nr}} \equiv 2\pi/\lambda_{\text{nr}}$) diffraction occurs at normal incidence, $\kappa_0=0$ and $\theta_R=0$. We consider the range of wavelengths between $\lambda_{\text{nr}}$ and $\lambda_{\text{gz}}$; here and throughout the paper the wavelengths we refer to are vacuum wavelengths.

\begin{figure}[htb]
\includegraphics[scale=0.3]{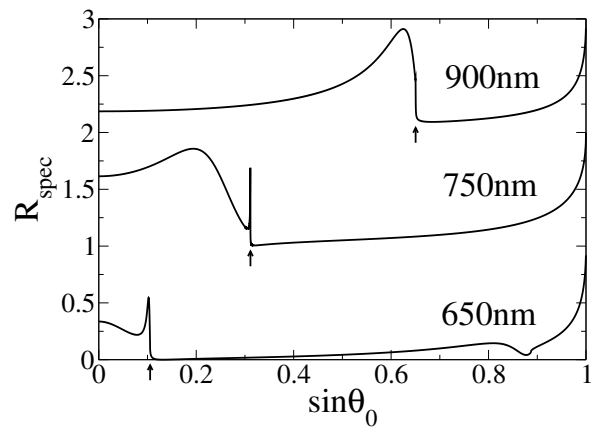}
\caption{Resonant structure of the specular reflection for s-polarized light and $\boldsymbol{\kappa}_0\propto \boldsymbol{\hat{\kappa}}_a$. Arrows identify the angles at which the Rayleigh line is crossed and diffraction appears. Plots are shifted by 1 to improve clarity. }
\label{frequencies}
\end{figure}

\begin{figure}[htb]
\includegraphics[scale=0.6]{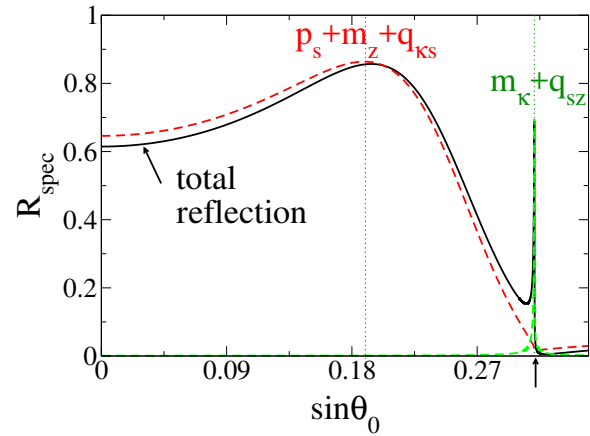}
\caption{Total specular reflection at $\lambda=750nm$ (continuous line) and the contributions from the two sets of multipole moments (dashed lines); the dotted lines show the SLMs supported by the the array. The arrow indicates an angle at which the Rayleigh line is crossed.}
\label{spec750}
\end{figure}

Although the intensity of the diffracted light shows no clear resonances in this range, the specularly reflected light exhibits one or two SLRs at angles of incidence less than the angle of
the Rayleigh line, $\theta<\theta_R$; see Fig.~\ref{frequencies}. 
To identify how the two independent sets of multipole moments (\ref{S1},\ref{S2}) contribute to the SLRs, we compare the contributions they make to the specular reflectivity. These are shown in Fig.~\ref{spec750} for a narrow angular scan at $\lambda=750nm$, 
and we see that each of the resonances can be associated with one of two sets of moments: a broad peak is associated with the resonance for the multipole moments $(p_s,m_z,q_{\kappa s})$ given by eq. (\ref{S1}), while the very narrow peak is associated with the higher order multipoles $(m_{\kappa},q_{sz})$ given by eq. (\ref{S2}). We shall see
that, as expected, the electric dipole makes the largest contribution to the first, and so we call it the \textit{dipolar} SLR; the second we call the \textit{multipolar} SLR. The resonant peaks at other wavelengths can also be clearly identified with the sets of multipoles:  the peak at $650nm$, for example, is a  multipolar SLR while the broad peak at $900nm$ is a dipolar SLR.   

\begin{figure}[htb]
\includegraphics[scale=0.6]{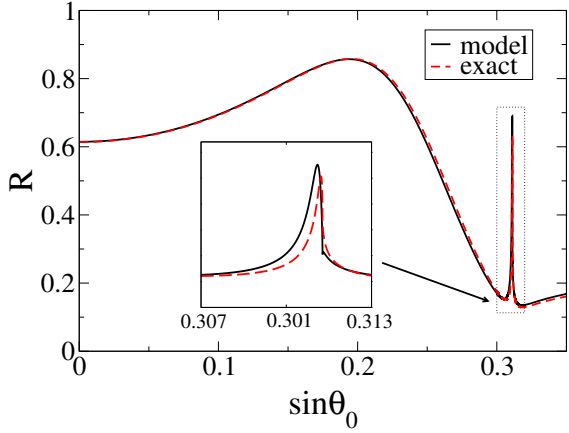}
\caption{Comparison of the total reflectivity predicted by the multipolar model, which includes multipole contributions from the electric and magnetic dipoles and the electric quadrupole, with the exact result obtained from the layered-multipole scattering method. }
\label{exact}
\end{figure}

Before considering the physics of these resonances, we confirm that the multipolar model (\ref{eq1}-\ref{eq2}) describes the optical response of the array correctly. We have compared the multipolar model results with an essentially exact numerical treatment, obtained using the layer-multipole scattering method,\cite{Stefanou2000} for wavelengths in the range $\lambda_{\mathrm{nr}}$ to $\lambda_{\mathrm{gz}}$ and with scans over all incident angles. We find that our multipolar model captures all the resonances that appear in the exact numerical calculation. To show the degree of agreement in describing each of the resonances, in Fig.~\ref{exact} 
we compare the multipolar model result at $\lambda=750nm$ with an  exact  treatment.\cite{Stefanou2000} Note that the multipolar model gives an excellent description of the dipolar SLRs associated with the set of equations (\ref{S1}), and a good description of the multipolar SLRs associated with the set of equations (\ref{S2}). At large wavelengths -- $\lambda\approx 805nm$ within the multipolar model and $\lambda\approx 785nm$ in the exact calculation -- the multipolar SLR merges with the Rayleigh anomaly and we observe a kink in reflectivity at $\theta = \theta_R$ rather than a SLR. The model thus correctly predicts the disappearance of the SLR, but there are some discrepancies between the exact calculation and the multipolar model in the description of set (\ref{S2}) close to the Rayleigh anomaly at long wavelengths. Nevertheless, even close to the anomaly the model correctly identifies the leading correction to each SLR beyond the simplest description with each SLR approximated by a response from one dominant multipole only. This point we discuss in detail below.

With the validity of truncating the multipole expansion at the level of the magnetic dipole and electric quadrupole confirmed for $\lambda\lesssim 800nm$, we now investigate whether the description can be further simplified by approximating each set of multipole moments (\ref{S1},\ref{S2}) by one dominant moment. Taking into consideration the dominant character of the electric dipole at long wavelengths, we might expect the electric dipole to make the dominant contribution to the resonances associated with the set of moments (\ref{S1}). Similarly, because of the negligible magnetic dipole response of an isolated sphere, we might expect that the inclusion of the magnetic dipole in the set (\ref{S2}) would lead to negligible effects. We shall see below that the first of these expectations is confirmed over a wide range of wavelengths, but the second is not. 

\begin{figure}[htb]
\includegraphics[scale=0.3]{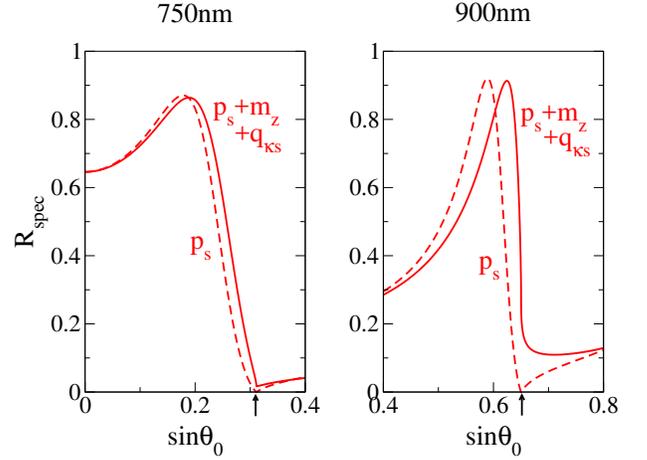}
\caption{The contribution to specular reflection from the set (\ref{S1}), compared with the contribution from the electric dipole only. Arrows identify the angles at which the Rayleigh line is crossed and diffraction appears.}
\label{sens_set1}
\end{figure}

We compare the total contribution to the specular reflection from the $(p_s,m_z,q_{\kappa s})$ set with the reflection calculated within the electric dipole approximation; see Fig.~\ref{sens_set1}. 
We see that the dipole approximation is accurate for broad peaks observed at small wavelengths. At large wavelengths, $\lambda\approx 900nm$, we observe that the inclusion of higher multipoles leads to some shift of the resonance towards the Rayleigh anomaly and to a non-zero reflectivity at $\theta=\theta_R$; we will refer to the non-zero value of the reflectivity at the onset of diffraction as the  \textit{suppression} of the anomaly. Finally, we note 
that the range of wavelengths over which the resonance is observed is only weakly affected by the coupling of the dipole to the higher multipoles. Both within the
full multipolar model and within the dipole approximation the SLR disappears at a cut-off wavelength $\lambda_{\text{cf}}$ close to wavelength at which the Rayleigh line moves to the right of the light line in \mbox{Fig.~\ref{ldisp},} $\lambda_{\text{cf}}\approx\lambda_{\text{gz}}$; the difference in $\lambda_{\text{cf}}$ within both approximations is of the order of few nanometers. 

\begin{figure}[htb]
\includegraphics[scale=0.3]{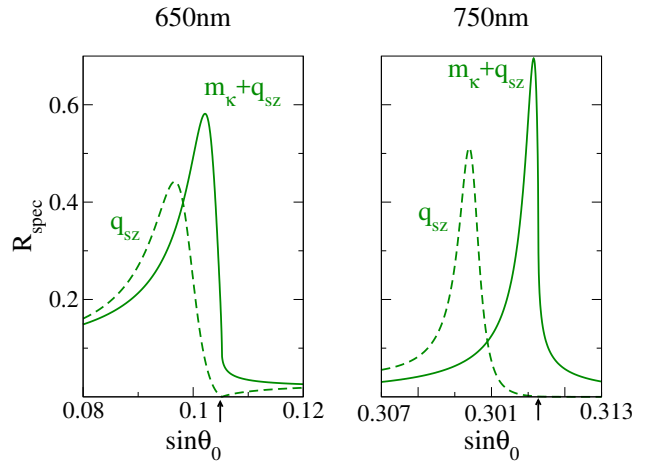}
\caption{The contribution to specular reflection from the set (\ref{S2}), compared with the contribution from the electric quadrupole only. Arrows indicate the angle at which the Rayleigh line is crossed and diffraction appears.}
\label{sens_set2}
\end{figure}

Next we compare the contribution to the specular reflection from the set $(m_{\kappa},q_{sz})$ with that of the electric quadrupole $q_{sz}$ only, see \mbox{Fig.~\ref{sens_set2}.} 
Again we observe that the inclusion of the magnetic response leads to a shift of the resonance and to the suppression of the Rayleigh anomaly, but this time the changes are much more pronounced and are significant over the whole range of wavelengths we consider. We also note that the coupling to the magnetic dipole significantly affects the wavelengths over which the multipolar SLR is observed; if the magnetic response of the sphere were neglected, the merging of the SLR with the Rayleigh anomaly would not occur and the SLR would be found for all wavelengths $\lambda \lesssim \lambda_{\text{gz}}$. 

Thus we see that even when there is a clearly dominant multipole moment, as for the set $(p_s, m_z, q_{\kappa s})$, the inclusion of other moments can lead to a change in the shape of the SLR and the suppression of the Rayleigh anomaly. And for the set $(m_{\kappa}, q_{sz})$, where one might expect the quadrupole moment to be dominant because of the small and non-resonant response of the magnetic dipole moment of an isolated sphere at these wavelengths, the inclusion of the magnetic dipole not only suppresses the Rayleigh anomaly, but significantly modifies the shape and the position of the SLR. To understand these features we first turn to a characterization of the SLRs as the excitation of effective  modes of the array. 

\section{Surface-lattice modes}
\label{normal modes}

By way of background, consider first a simple planar system, such as a thin metal film surrounded by a dielectric with index of refraction $n$. Over a range of frequencies the metal film will support long-range surface plasmons.\cite{Berini,SPP} For a real frequency $\omega$ each possible long-range surface plasmon is characterized by a wave vector $\boldsymbol{\kappa}$ in the plane of the film. Because of absorption in the metal the plasmons decay as they propagate, and the corresponding wave vector $\boldsymbol{\kappa}$ is complex. The complex  $\boldsymbol{\kappa}$ identifies a pole in the appropriate response coefficient of the structure, which very generally can be taken to identify the modes of a system. If absorption is neglected, $\boldsymbol{\kappa}$ moves onto the real axis. But since the fields are evanescent in the dielectric, $|\boldsymbol{\kappa}|> \tilde{\omega}n$, the excitation cannot be accessed by an incident plane wave, for which $|\boldsymbol{\kappa}_0|<\tilde{\omega}n$. In practice, one way to couple into such an excitation is with the aid of a grating structure.\cite{Berini,SPP} If the grating is characterized by a wave vector $\boldsymbol{K}$ such that $\boldsymbol{\kappa} \approx \boldsymbol{\kappa}_0+\boldsymbol{K}$, coupling can be achieved.  

Returning to our lattice of spheres, we analyze SLRs by considering the lattice as an effective planar structure plus a grating. With the neglect of loss, the SLRs observed in the specular reflection correspond to the coupling into modes of the structure, each characterized by a wave vector $\boldsymbol{\kappa}$ and using one of the $\boldsymbol{K}$ of the lattice such that $|\boldsymbol{\kappa}_0+\boldsymbol{K}|=\kappa$.\cite{Carron1986, Ohtaka1982, Bendana2009, Campione2014}. The strong coupling into the mode and back into the specularly reflected light leads to the large reflectivity. We refer to these modes as surface-lattice modes (SLMs). No evidence of SLM appears in the diffracted intensity, of course, since $\kappa>\tilde{\omega}n$ and at $\boldsymbol{\kappa}$ there is no radiation into the dielectric.

The loss relevant here is due both to absorption in the metal and to coupling through $-\boldsymbol{K}$ of the SLM back to the incident wave vector $\boldsymbol{\kappa}_0$; the light at $\boldsymbol{\kappa}_0+\boldsymbol{K}$ coupled back at $\boldsymbol{\kappa}_0$ contributes to a loss of the SLM through the increased specular reflectivity that signals its very presence. So to identify a particular SLM in the lossless limit we need to neglect the elements of the response associated with the absorption in the metal
and with the radiation at $\boldsymbol{\kappa}_0$, identify the response to the incident field of the relevant combination of multipoles, and for each frequency $\omega$ of interest identify the $\boldsymbol{\kappa}$ at which the response diverges. The resulting function $\omega(\boldsymbol{\kappa})$ identifies the dispersion relation of the SLM in the lossless limit, what we call the
``ideal" dispersion relation.

\subsection{Energy balance}
To construct these ideal dispersion relations the first step is then to isolate the loss mechanisms. To clarify the physics we first consider the energy balance in an array of dipoles. Within the dipole approximation, the response of an electric dipole to the incident field can be described in terms of an effective polarizability of the array,\cite{DeAbajo2007,Zou2004,Zou2004b} $\boldsymbol{p} = \ovb{\alpha}^{\text{eff}} \boldsymbol{E}^{\text{inc}}$, where
\begin{align}
\left(\ovb{\alpha}^{\text{eff}} \right)^{-1} = \ovb{\alpha}^{-1} -\ovb{\mathcal{G}}^{\text{Ep}},\label{eff}
\end{align}
and where $\ovb{\alpha} = \alpha^{pE}\ovb{U}$. The real part of the inverse effective polarizability (\ref{eff}) does not affect the energy balance in the system.\cite{Belov2005,paper2} The imaginary part of the inverse  polarizability of the sphere, $\ovb{\alpha}^{-1}$, determines the resonance broadening due to the absorption losses. The imaginary part of the periodic Green function identifies radiation loss and is known in a closed form,\cite{Belov2005,paper2}  
\begin{align}
\text{Im}\ovb{\mathcal{G}}^{\text{Ep}}  = & \sum_{\boldsymbol{n}\forall |\boldsymbol{\kappa}_{\boldsymbol{n}}|<\tilde{\omega}n} \text{Im}\ovb{\mathcal{G}}^{\text{Ep}}_{\boldsymbol{n}},\label{imag}
\end{align} 
where 
\begin{align*}
\text{Im}\ovb{\mathcal{G}}^{\text{Ep}}_{\boldsymbol{n}} = \frac{ w_{\boldsymbol{n}}^2 \boldsymbol{\hat{\kappa}}_{\boldsymbol{n}} \boldsymbol{\hat{\kappa}}_{\boldsymbol{n}} +(\tilde{\omega}n)^2 \boldsymbol{\hat{s}}_{\boldsymbol{n}} \boldsymbol{\hat{s}}_{\boldsymbol{n}} +\kappa^2_{\boldsymbol{n}} \boldsymbol{\hat{z}}\boldsymbol{\hat{z}}}{2A_c \epsilon_0 n^2 w_{\boldsymbol{n}}},
\end{align*}
and we have put $\boldsymbol{\hat{\kappa}}_{\boldsymbol{n}} = \boldsymbol{\kappa}_{\boldsymbol{n}}/|\boldsymbol{\kappa}_{\boldsymbol{n}} | $, $\boldsymbol{\hat{s}}_{\boldsymbol{n}}=\boldsymbol{\hat{\kappa}}_{\boldsymbol{n}} \times\boldsymbol{\hat{z}}$. We note that each of the terms  $\text{Im}\ovb{\mathcal{G}}^{\text{Ep}}_{\boldsymbol{n}}$ in the sum (\ref{imag}) accounts for the radiation reaction associated with a diffracted beam radiating at an angle $\theta_{\boldsymbol{n}} = \text{sin}^{-1} \left[|\boldsymbol{\kappa}_{\boldsymbol{n}}|/(\tilde{\omega}n)\right]$; see eqs. (\ref{Rn},\ref{Tn}).

These considerations generalize immediately to an array described by a full set of multipole moments (\ref{eq1}-\ref{eq2}). As in the dipole example, the imaginary part of the inverse proper polarizabilities determine the Ohmic losses. The radiative losses are determined by the imaginary part of the periodic Green functions $\ovb{\mathcal{G}}^{\text{Ep}}, \ovb{\mathcal{G}}^{\text{Bm}}, \ovb{\mathcal{G}}^{\text{Fq}},
\ovb{\mathcal{G}}^{\text{Em}},
\ovb{\mathcal{G}}^{\text{Bp}}
$, and by the real part of the Green functions
$\ovb{\mathcal{G}}^{\text{Eq}}, \ovb{\mathcal{G}}^{\text{Fp}}, \ovb{\mathcal{G}}^{\text{Bq}},
\ovb{\mathcal{G}}^{\text{Fm}}$.\cite{paper2} The contributions to the periodic Green functions that describe radiative losses are known in  closed form,\cite{paper2} and in the pattern of (\ref{imag}) are given by a Fourier sum over wave vectors $\boldsymbol{\kappa}_{\boldsymbol{n}}$, $|\boldsymbol{\kappa}_{\boldsymbol{n}}|<\tilde{\omega}n$, with the Fourier component at $\boldsymbol{\kappa}_{\boldsymbol{n}}$ accounting for the radiation reaction associated with the beam radiating at an angle $\theta_{\boldsymbol{n}}$.

\subsection{Dispersion relations}
With the channels for energy loss due to both absorption and radiation
identified, we are in a position to find the SLMs of the array in the lossless limit. For the illumination conditions discussed in section III, the matrices $\mathbb{S}$ and $\mathbb{S}'$, eqs. (\ref{S1}-\ref{S2}), describe the response as a function of $\boldsymbol{\kappa}_0$, the in-plane component of the incident wave vector. The divergences in the response, and thus the modes, are associated with the vanishing of the determinants of the matrices in the limit of no loss. We now neglect absorption and radiation by dropping the imaginary parts of the inverse proper polarizabilities and the parts of the relevant Green functions associated with radiation reaction. Writing the determinants of $\mathbb{S}$ and $\mathbb{S}'$ in this limit as $d(\boldsymbol{\kappa}_0)$ and $d'(\boldsymbol{\kappa}_0)$ respectively, we find that these vanish for certain $\boldsymbol{\kappa}_0$. In general $d(\boldsymbol{\kappa}_0)$ and $d'(\boldsymbol{\kappa}_0)$ are periodic functions over the Brillouin zone, so any vanishing of determinant for $\boldsymbol{\kappa}_0$, $|\boldsymbol{\kappa}_0|<\tilde{\omega}n$, is replicated with the periodicity of the Brillouin zone over all reciprocal space; thus each $\boldsymbol{\kappa}_0$ we associate with a SLM at $\boldsymbol{\kappa}_{\boldsymbol{n}} = \boldsymbol{\kappa}_0 + \boldsymbol{K}_{\boldsymbol{n}}$, $|\boldsymbol{\kappa}_{\boldsymbol{n}}|>\tilde{\omega}n$. Due to periodicity of the system there is an ambiguity in the choice of $\boldsymbol{K}_{\boldsymbol{n}}$ and thus in the choice of $\boldsymbol{\kappa}_{\boldsymbol{n}}$. Out of all the possible reciprocal vectors we choose the $\boldsymbol{K}_{\boldsymbol{n}}$ that corresponds to $\boldsymbol{\kappa}_{\boldsymbol{n}}$ closest to the light line, which for the parameter space that we consider corresponds to $\boldsymbol{\kappa}_{\boldsymbol{n}}$ in the second Brillouin zone. We now claim that even when absorption and radiation are included, the SLRs can be understood as arising from a coupling of the incident field at $\boldsymbol{\kappa}_0$ to the SLM at $\boldsymbol{\kappa}_{\boldsymbol{n}}$ and then coupling back to $\boldsymbol{\kappa}_0$ by the vector $-\boldsymbol{K}_{\boldsymbol{n}}$. In Figure \ref{spec750} we plot in dotted lines the location of the divergence in $d(\boldsymbol{\kappa}_0)$ that we find (associated
with the set of moments $(p_s,m_z,q_{\kappa s})$), and the location of the divergence
in $d'(\boldsymbol{\kappa}_0)$ that we find (associated with the set of moments $(m_\kappa, q_{sz})$). We
see that there is excellent agreement between the location of the divergences in the lossless limit and the peaks that appear in the full calculation, both for the scenario depicted in Fig.~\ref{spec750} and at other wavelengths as well, justifying the physical picture that we presented.

\begin{figure}[htb]
\includegraphics[scale=0.3]{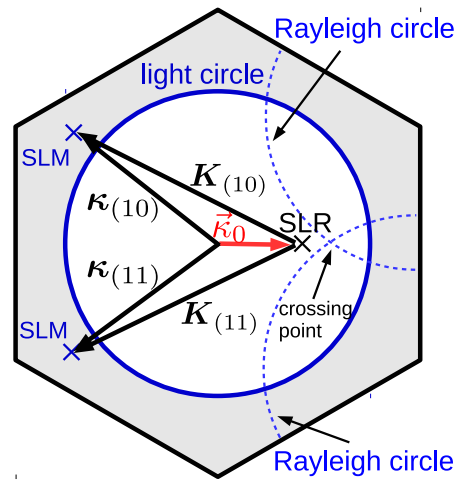}
\caption{For a fixed $\omega$, the SLMs at $\boldsymbol{\kappa}_{(10)} = \boldsymbol{\kappa}_0 + \boldsymbol{K}_{(10)}$ and $\boldsymbol{\kappa}_{(11)} = \boldsymbol{\kappa}_0 + \boldsymbol{K}_{(11)}$. The solid blue line indicates the ``light circle", the set of $\boldsymbol{\kappa}$ for which at chosen frequency $\kappa = n\omega/c$. The dotted blue lines indicate the ``Rayleigh circles", which are light circles centered at reciprocal lattice sites other than that at the origin; diffraction associated with a particular $\boldsymbol{K}$ occurs when $\boldsymbol{\kappa}_0$ crosses its Rayleigh circle.}
\label{modes}
\end{figure}

\begin{figure}[htb]
\includegraphics[scale=0.3]{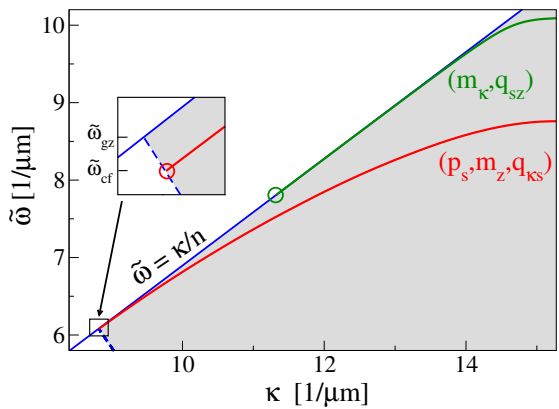}
\caption{Ideal dispersion relations of the two SLMs. Blue lines identify the light line (solid) and the Rayleigh line (dashed) of the SLM. The terminations of the dispersion relations are indicated by circles.}
\label{disp}
\end{figure}

Because of a high symmetry of the illumination conditions discussed in section III, the value of $\boldsymbol{\kappa}_0$ within the light line where a divergence in $d(\boldsymbol{\kappa}_0)$ or $d'(\boldsymbol{\kappa}_0)$ occurs is in each case associated with two $\boldsymbol{\kappa}$ of equal magnitude $\kappa$, $\boldsymbol{\kappa}_{(10)} = \boldsymbol{\kappa}_0 + \boldsymbol{K}_{(10)}$ and
$\boldsymbol{\kappa}_{(11)} = \boldsymbol{\kappa}_0 + \boldsymbol{K}_{(11)}$, and thus two SLMs. The scenario is sketched in reciprocal space at a fixed $\omega$ in Fig.~\ref{modes}. 
The light circle is the set of $\boldsymbol{\kappa}$ for which $\kappa = \tilde{\omega}n$; the point on the light circle crossed when moving in the direction of $\boldsymbol{\hat{\kappa}}_0$ identifies the point on the light line of Fig.~\ref{ldisp} at the chosen frequency. The Rayleigh circles are the circles of the same radius $\tilde{\omega}n$ centered at the other reciprocal lattice sites. It is easy to confirm that the light circle and the Rayleigh circles identify the curves in reciprocal space where the periodic Green functions undergo singular and, more generally, non-analytic behavior, as discussed in section IIB. In Fig.~\ref{modes}, the first point on a Rayleigh circle that is crossed when moving through reciprocal space in the direction of $\boldsymbol{\hat{\kappa}}_0$  identifies the point on the Rayleigh line of Fig.~\ref{ldisp} at the chosen frequency. This point is identified in Fig.~\ref{modes} as the ``crossing point" and is associated with (10) and (11) diffraction orders due to the symmetry. The SLRs found in the vicinity of the ``crossing point" are associated with the SLMs that would result were loss ignored; those modes reside beyond the light circle as is indicated. For different frequencies $\omega$  the value of $\kappa = |\boldsymbol{\kappa}_{(10)}| = |\boldsymbol{\kappa}_{(11)}|$ associated with the SLMs is different, and we can plot out the dispersion relation $\tilde{\omega}(\kappa)$. This is presented in \mbox{Fig.~\ref{disp},} 
both for the SLM associated with the dipolar SLR, and that associated with the multipolar SLR.

The dispersion relations are what one would expect for those characterizing excitations bound to a surface or thin film, in that at lower wave numbers they are close to the light line, and they pull away at higher wave numbers. What is novel is their termination. The multipolar SLM terminates at a finite frequency on the light line, while the dipolar SLM terminates at a lower frequency away from the light line.

\begin{figure}[htb]
\includegraphics[scale=0.3]{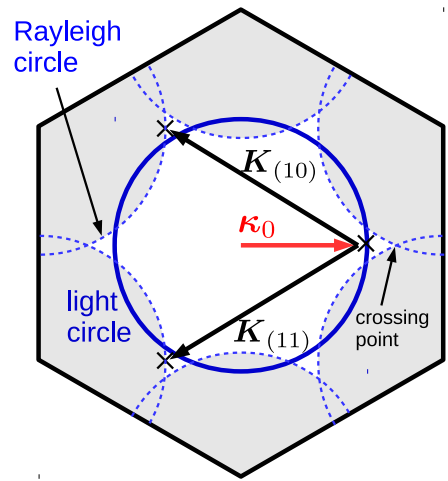}
\caption{Scenario in the reciprocal lattice at the cut-off frequency $\omega_{\text{cf}}$. The SLM wave vector $\boldsymbol{\kappa} = \tilde{\omega}_{\text{cf}}n\boldsymbol{\hat{\kappa}}_a+\boldsymbol{K}$, for $\boldsymbol{K}=\boldsymbol{K}_{(10)}$ or $\boldsymbol{K}=\boldsymbol{K}_{(11)}$, resides on a Rayleigh circle.}
\label{restricted}
\end{figure}

We can understand the termination of the dipolar SLM 
with the aid of the shaded region in Fig.~\ref{disp}. At each $\omega$ it identifies the magnitudes $\kappa=|\boldsymbol{\kappa}|$ of the wave vectors $\boldsymbol{\kappa}$, which (a) are of the form $\boldsymbol{\kappa} = \kappa_0\boldsymbol{\hat{\kappa}}_a  + \boldsymbol{K}$
where $\boldsymbol{K} = \boldsymbol{K}_{(10)}$ or $\boldsymbol{K}=\boldsymbol{K}_{(11)}$ (b) are beyond the light line but within the second Brillouin zone, and (c) for which the periodic Green functions remain analytic, which is to say that the light circle and the Rayleigh circles are avoided. In general, the wave vectors on the Rayleigh circles can be excited with or without the help of the grating. The wave vectors on the Rayleigh circle that can be excited without the help of the grating are of the form $\boldsymbol{\kappa} = \boldsymbol{\kappa}_0$, for example the crossing point in Fig.~\ref{modes}, and they identify a point on the Rayleigh line of Fig.~\ref{ldisp}. The wave vectors on the Rayleigh circles that are excited with the help of the grating are of the form $\boldsymbol{\kappa} =\tilde{\omega}n\boldsymbol{\hat{\kappa}}_a+\boldsymbol{K}$, where $\boldsymbol{K} = \boldsymbol{K}_{(10)}$ or $\boldsymbol{K} = \boldsymbol{K}_{(11)}$; see for example Fig.~\ref{restricted}, drawn for a lower frequency than Fig.~\ref{modes}. These wave vectors identify the onset of diffraction, and can thus be taken to define a new Rayleigh line that we plot in Fig.~\ref{disp} as a dashed line; due to the symmetry at each frequency there are two such wave vectors of equal magnitude.  In consideration of the SLMs it is the Rayleigh line of Fig.~\ref{disp} that is important. Below $\tilde{\omega}_{\text{gz}}$ this Rayleigh line restricts the shaded area in Fig.~\ref{disp}. When the dispersion relation of the SLM meets this Rayleigh line, at $\tilde{\omega}_{\text{cf}}$, the singular change in the Green functions terminates the SLM. The situation in reciprocal space at this frequency is depicted in Fig.~\ref{restricted},
where each SLM falls precisely on the Rayleigh circle of a reciprocal lattice vector. We note that were the magnetic dipole
neglected there would be a small change of the dipolar dispersion relation, resulting in a small change in the cut-off frequency; however, the underlying description of the SLM termination would remain the same.

The inclusion of magnetic response in the description of multipolar SLM, on the other hand, changes the cut-off frequency from a value just below  $\omega_{\text{gz}}$ to a value significantly above it; see Fig.~\ref{disp}. This dramatic change in cut-off frequency of the multipolar SLM cannot be identified with simple features of the reciprocal lattice geometry, as could the location of the cut-off frequency of the dipolar SLM. To understand the termination of the multipolar SLM, for which the coupling between the magnetic dipole and electric quadrupole terms plays an important role, and to understand in more detail how the singular change in the Green function terminates the dipolar SLM, it is useful to develop a simplified model for the SLMs based on expansions of
the periodic Green functions.

\section{Simplified model}
\label{simple model}

With the neglect of loss due to absorption or diffraction, each SLM we have identified involves a set of wave vectors, each wave vector in the set differing from another in the set by a translation over a reciprocal lattice. For much of the parameter space considered in the last section, at least one wave vector in the set lies close to the light line, and we considered the dispersion relation of the SLM associated with it. In the vicinity of the light line periodic Green functions can diverge, and it is this divergent behaviour that plays a dominant role in how the dispersion relation of the SLM is established. In this section we establish a simplified model for the SLMs based on isolating the divergent contribution.

We begin by considering the dispersion relation of the dipolar SLM. For simplicity we neglect the contribution to the dipolar SLM from the electric quadrupole moment of the spheres, as the corrections beyond the electric dipole approximation come mainly from the magnetic response. In the absence of the quadrupole terms  eq. (\ref{S1}) reduces to
\begin{align}
\mathbb{S}_{\text{pm}} \left( \begin{array}{c}
p_s \\
m_z
\end{array}
 \right) = \left( \begin{array}{c}
 E_{\text{inc},s} \\
 B_{\text{inc},z}
\end{array}  
\right),\label{Seq}
\end{align}
where the matrix $\mathbb{S}_{\text{pm}}$ is given by,
\begin{align}
\mathbb{S}_{\text{pm}} = \left( \begin{array}{cc}
\left( \alpha^{\text{pE}}  \right)^{-1} -\mathcal{G}^{\text{EP}}_{ss} & -\mathcal{G}^{\text{EM}}_{sz} \\
-\mathcal{G}^{\text{BP}}_{zs}  & \left( \alpha^{\text{mB}}  \right)^{-1} -\mathcal{G}^{\text{BM}}_{zz}
\end{array} \right).\label{Spm}
\end{align} 
and the components of the dyadics and vectors are written in the $(\boldsymbol{\hat{s}},\boldsymbol{\hat{\kappa}},\boldsymbol{\hat{z}})$ basis associated with the incident light, that is here $\boldsymbol{\hat{\kappa}} = \boldsymbol{\hat{\kappa}}_0$, etc. To evaluate the periodic Green functions in an approximate way that respects the important physics, we note that each periodic Green function has a contribution from a Fourier sum that includes components at each $\boldsymbol{\kappa}_{\boldsymbol{n}} = \boldsymbol{\kappa}_0 + \boldsymbol{K}$. Each of these components can diverge as $w_{\boldsymbol{n}}^{-1} = \left((\tilde{\omega}n)^2 - \kappa^2_{\boldsymbol{n}}\right)^{-1/2}$ for $\boldsymbol{\kappa}_{\boldsymbol{n}}$ close to the light-line. Looking at the scenario shown in Fig.~\ref{modes}, for $\boldsymbol{\kappa}_0$ close to the SLR we have $\boldsymbol{\kappa}_{(10)}$ and $\boldsymbol{\kappa}_{(11)}$
close to the light-line; we treat these Fourier components separately. Due to symmetry $w_{(10)} = w_{(11)} \equiv w_{R}$, where $w_R$ is a function of two independent variables $\boldsymbol{\kappa}_0$ and $\tilde{\omega}$, and the Fourier components at $\boldsymbol{\kappa}_{(10)}$ and $\boldsymbol{\kappa}_{(11)}$ diverge as $w_R^{-1}$; the next term in an expansion of these Fourier components in powers of $w_R$ is of order $w_R$, and we neglect
it. The remaining Fourier components and the contribution from the multipoles at the origin are analytic in $w_R$ as $w_R \rightarrow 0$, and the leading non-vanishing terms are of order $w_R^0$; we keep only this order and denote the sum of all such terms in the periodic Green functions as $\tilde{\mathcal{G}}^{\text{EP}}_{ss}$, $\tilde{\mathcal{G}}^{\text{EM}}_{sz}$, etc. We then use these approximate expressions to identify the singular and the leading analytic terms in $w_R$ as $w_R \rightarrow 0$ for each of the three independent elements of the
matrix $\mathbb{S}_{\text{pm}}$, eq. (\ref{Spm}), taken in the limit of no loss. The analytic contributions are given by the functions independent of $w_R$,
\begin{align}
\beta^{-1}_{p} &\equiv \text{Re}\left( \alpha^{\text{pE}}  \right)^{-1} - \tilde{\mathcal{G}}^{\text{EP}}_{ss},\label{bp} \\
\beta^{-1}_{m}  &\equiv   \text{Re}\left( \alpha^{\text{mB}}  \right)^{-1} - \tilde{\mathcal{G}}^{\text{BM}}_{zz}, \\
\beta^{-1}_{cp} &\equiv  \tilde{\mathcal{G}}^{\text{EM}}_{sz} = -\tilde{\mathcal{G}}^{\text{BP}}_{zs}.\label{bcp}
\end{align}
We denote the singular contribution to the electric dipole Green function as
\begin{equation}
\mathcal{G}_{\text{sng}} = \text{Re} \frac{i}{\epsilon_0 n^2} \frac{ (\tilde{\omega}n)^2 - K^2_{s} }{A_c w_R},\label{FLQ}
\end{equation} 
where $K_{s} = |\boldsymbol{K}_{(10)}\cdot \boldsymbol{\hat{s}}| = |\boldsymbol{K}_{(11)}\cdot \boldsymbol{\hat{s}}|$. The singular contributions to the other Green function terms differ from $\mathcal{G}_{\text{sng}}$ by a factor $\eta$,
\begin{displaymath}
\eta  = \frac{n}{c}\frac{\tilde{\omega}n}{\sqrt{(\tilde{\omega}n)^2 - K_{s}^2}}.
\end{displaymath}
We can treat the $\beta$'s  as ``dressed" polarizabilities of a sphere on a lattice in the absence of absorption and radiation losses, since they do not diverge at the onset of the Rayleigh anomalies and they depend on frequency only. However, even if the $\alpha^{pE}$ and $\alpha^{mB}$ are to good approximation Lorentzian functions of frequency, in general $\beta_p$, $\beta_m$, and of course $\beta_{\text{cp}}$ will not be Lorentzian, due to the frequency dependence of the Green function terms. Nevertheless, the frequency dependence of the $\beta$ coefficients identifies the SLM dispersion relation, as we now show.

SLMs are identified by wave vectors $\boldsymbol{\kappa}$ at which, with absorption and diffraction neglected, the determinant of the matrix (\ref{Spm}) vanishes. Within the dipole approximation the condition for SLM is given by
\begin{equation}
\beta_p \mathcal{G}_{\text{sng}} = 1. \label{deter1}
\end{equation}
When the coupling to magnetic dipole is included, the determinant of the matrix $\mathbb{S}_{\text{pm}}$ (here taken in the absence of loss) is quadratic in periodic Green functions. Nevertheless, a cancellation of terms that are quadratic in $\mathcal{G}_{\text{sng}}$ results in a condition for SLM of the same form as before,
\begin{equation}
\beta_{\text{pm}} \mathcal{G}_{\text{sng}} = 1, \label{deter2}
\end{equation}
but with a dressed polarizability renormalized by the coupling to magnetic dipole. Here $\beta^{-1}_{\text{pm}}  \equiv \beta^{-1}_p + \Delta_{\beta}$, where the correction due to the coupling is given by
\begin{align}
\Delta_{\beta} =  \frac{(\eta\beta^{-1}_p - \beta^{-1}_{\text{cp}})^2}{2\eta \beta^{-1}_{\text{cp}} - \eta^2\beta^{-1}_p - \beta^{-1}_m}.\label{Dsng}
\end{align}
Using eq. (\ref{FLQ}) in (\ref{deter1},\ref{deter2}) we solve for the value of $i^{-1} w_R$ that satisfies the SLM condition respectively in the dipole approximation and when the magnetic dipole is taken into account; the solution characterizes the distance of the SLM at $\boldsymbol{\kappa} = \boldsymbol{\kappa}_{(10)}$ and $\boldsymbol{\kappa} = \boldsymbol{\kappa}_{(11)}$ in each of the approximations from the light-line. We find that this quantity is proportional to the dressed polarizability $\beta$,
\begin{equation}
\frac{1}{i} w_R = \frac{1}{\epsilon_0} \frac{(\tilde{\omega}n)^2 - K_s^2}{n^2 A_c} \beta,\label{dipole}
\end{equation}
with and without the coupling to magnetic dipole included, where respectively $\beta=\beta_{\text{pm}}$ and $\beta=\beta_{\text{p}}$. From eq. (\ref{dipole}) it follows that the SLM merges with the light-line when  $\beta$ vanishes. 

\begin{figure}[htb]
\includegraphics[scale=0.3]{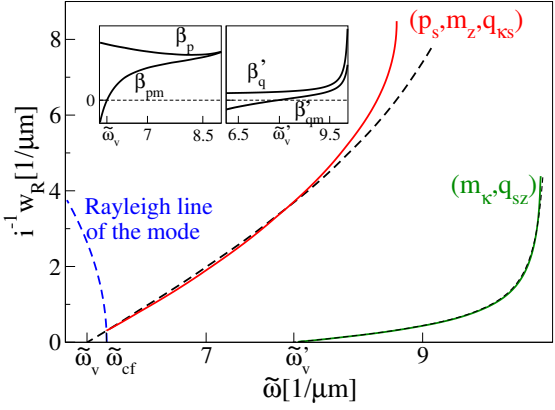}
\caption{Dispersion relations found within the multipolar model (continuous lines) and within the simplified model (dashed lines). Insets show the ``dressed" polarizabilities of the dipolar (left) and multipolar (right) SLMs with and without the coupling to magnetic dipole. When the coupling is included, the polarizabilities vanish at $\tilde{\omega}_v$ and $\tilde{\omega}'_v$ respectively.}
\label{smodel}
\end{figure}

We note if $\alpha^{\text{pE}}$ is well approximated by a Lorentizan, $\left(\alpha^{\text{pE}}\right)^{-1} \approx A(\omega_0^2-\omega^2)-i\Gamma$, the inverse dressed polarizability in the dipole approximation with loss neglected is of the form
\begin{equation}
\beta^{-1}_p  = A( \omega_0^2 - \omega^2)  - \tilde{\mathcal{G}}^{\text{EP}}_{ss}.  \label{betap}
\end{equation}
The resulting polarizability $\beta_p$ does not vanish, and so from (\ref{dipole}) we see there is a finite distance
of the SLM to the light line within the dipole approximation, and no
termination of the SLM. However, when the coupling to the magnetic
dipole is included, the polarizability $\beta_{\text{pm}}$ vanishes when the correction (\ref{Dsng}) diverges. This is confirmed by the plots in Fig.~\ref{smodel}
for our system, where indeed $\beta_{\text{p}}$ is finite while $\beta_{\text{pm}}$ vanishes at a frequency we label $\omega_v$; see the inset. As the frequency is lowered, one might then expect that at $\omega_v$ the normal SLM dispersion relation would reach the light line and the SLM would terminate. 

Yet this scenario is pre-empted
by the fact that at a frequency $\omega_{\text{cf}}>\omega_v$ the position of the SLM reaches the Rayleigh circle, as discussed above (see Fig.~\ref{restricted}), and as the Rayleigh
circle is approached a new singular contribution to the periodic Green functions becomes important. When loss and the radiative
contributions to the periodic Green function are ignored, this term vanishes until the Rayleigh circle is crossed, but it gives a divergent contribution at the crossing. This additional singular contribution to the Green function results in the vanishing of the SLM, which beyond the Rayleigh circle is no longer described by the simplified dispersion relations presented above, where only one singular term was taken into account.

The situation for the multipolar SLM is completely different, although a similar analysis can be applied. We introduce ``dressed" polarizabilties that describe the leading analytic contributions to the three independent elements of matrix $\mathbb{S}'$ taken in the limit of no loss,
\begin{align}
\beta^{\prime -1}_{\text{q}} &\equiv \text{Re}\left(\alpha^{\text{qF}}\right)^{-1}- 2\tilde{\mathcal{G}}^{\text{Fq}}_{szsz}, \nonumber\\
\beta^{\prime -1}_{\text{m}} &\equiv \text{Re}\left( \alpha^{\text{mB}} \right)^{-1} -\tilde{\mathcal{G}}^{\text{Bm}}_{\kappa\kappa}, \nonumber\\
\beta^{\prime -1}_{\text{cp}} &\equiv \tilde{\mathcal{G}}^{\text{Bq}}_{\kappa s z} = -\tilde{\mathcal{G}}^{\text{Fm}}_{sz \kappa},
\end{align}
where $\tilde{\mathcal{G}}^{\text{Fq}}_{szsz}$,  $\tilde{\mathcal{G}}^{\text{Bm}}_{\kappa \kappa}$, etc. denote the leading analytic contributions to periodic Green functions. The singular contribution to the quadrupole Green function we denote as
\begin{equation}
\mathcal{G}'_{\text{sng}} = \text{Re}\frac{i}{\epsilon_0 n^2} \frac{(\tilde{\omega}n)^2 K_{s}^2 }{4A_c w_R},
\end{equation}
with the singular contributions to other Green functions differing by a parameter $\eta' = 1/(\tilde{\omega}c)$. In this notation the condition for the SLMs is given by
\begin{equation}
\frac{1}{i}w_R = \frac{1}{\epsilon_0 n^2} \frac{(\tilde{\omega}n)^2 K_{s}^2}{2 A_c}\beta',\label{multipolar}
\end{equation}
where $\beta' = \beta'_{\text{q}}$ when the coupling to magnetic dipole is neglected, and $\beta' = \beta'_{\text{qm}}$ when magnetic dipole is in included in the description; again, $i^{-1}w_R$ characterizes the distance of the SLM dispersion relation from the light line. The quadrupole dressed polarizability renormalized by the coupling is given by
\begin{equation}
\beta^{\prime -1}_{\text{qm}} = \beta^{\prime -1}_{\text{q}} +\Delta'_{\text{qm}},
\end{equation}
with the correction
\begin{equation}
\Delta'_{\text{qm}} = \frac{\left(\eta^{\prime}\beta^{\prime -1}_{\text{q}} - \beta^{\prime -1}_{\text{cp}}  \right)^2}{2\eta^{\prime}\beta^{\prime -1}_{\text{cp}}-\eta^{\prime 2}\beta^{\prime -1}_{\text{q}} - \frac{1}{2}\beta^{\prime -1}_{\text{m}} }.
\end{equation}
As in our discussion of the dipolar SLM, we find
that when the SLM is approximated by one dominant multipole (here the
electric quadrupole), the dressed polarizability does not vanish, $\beta'_{\text{q}}\neq 0$; thus were only this multipole present there would be no termination of the SLM dispersion. However, when the coupling to the magnetic dipole is included the renormalized dressed polarizability $\beta'_{\text{qm}}$ vanishes at a frequency $\omega'_v$; here $\omega'_v>\omega_{\text{cf}}$, and so the merging of the SLM dispersion relation with the light light is not pre-empted by the presence of a Rayleigh circle, and the dispersion relation terminates at $\omega'_v$, as shown by the green circle in Fig.~\ref{disp}.

We emphasize that the frequency at which the multipolar SLM vanishes, $\omega'_v$, strongly depends on all the details of the system, including both the lattice geometry and the polarizabilities of the spheres. This is in contrast to the situation for the dipolar SLM, which terminates before its dispersion relation reaches the light line. There the frequency at which the SLM vanishes, $\omega_{\text{cf}}$, depends mostly on the geometry of the reciprocal lattice which determines the Rayleigh anomaly line associated with the SLM of Fig.~\ref{disp}.

\section{Surface-lattice modes and the sensitivity of SLRs to non-resonant multipoles}
\label{discussion}

We have shown that the SLM dispersion relations identify the position of the SLRs and the range of frequencies over which they exist. Now we show that the sensitivity of the SLRs to the inclusion of non-resonant magnetic dipole in their description can be linked to the dispersion relations as well. The link we provide is only approximate, but nevertheless it gives an insight into a large sensitivity of the multipolar SLRs and of the dipolar SLRs at long wavelengths.

Besides their position, the SLRs are also characterized
by their widths, which are mostly restricted by the rapid variations of the reflection close to the Rayleigh line; see Figs. \ref{sens_set1}, \ref{sens_set2}. For SLRs that are well described with the electric dipole approximation the reflection vanishes at the Rayleigh line, and the distance in reciprocal space from the peak to the Rayleigh line can be taken as an indication of the width.\cite{Markel2005} When the contribution from more than one multipole becomes important the reflection at the Rayleigh line is in general finite, but it decreases rapidly as $\boldsymbol{\kappa}_0$ moves beyond the line. Thus quite generally we can take the reciprocal space distance $\Gamma = \left|\boldsymbol{\kappa}_\text{R} - \boldsymbol{\kappa}_{\text{SLR}}\right|$, where $\boldsymbol{\kappa}_{\text{SLR}}$ is the wave vector of an SLR and $\boldsymbol{\kappa}_\text{R}$ is the wave vector of the Rayleigh line, as an indicator of the width of the SLR. Even though the distance  $\Gamma$ gives only an approximate characterization of the SLR profile, it has an advantage over more sophisticated treatments in that it simultaneously identifies both the width and the position of an SLR. We refer to it as the ``width parameter", and a straight-forward calculation shows that it can be related to the wave vector of the SLM, $\boldsymbol{\kappa}_{\text{SLM}} = \boldsymbol{\kappa}_{\text{SLR}}+\boldsymbol{K}_{\boldsymbol{n}}$ with $\boldsymbol{n}=(1,0)$ or $\boldsymbol{n}=(1,1)$, by  
\begin{equation}
\Gamma =d\left[ \frac{\kappa_{\text{SLM}} + \tilde{\omega}n}{\sqrt{\kappa_{\text{SLM}}^2-K_s^2}+\sqrt{\left(\tilde{\omega}n\right)^2-K_s^2}} \right],\label{Gamma}
\end{equation} 
where $\kappa_{\text{SLM}} = \left| \boldsymbol{\kappa}_{\text{SLM}} \right|$, $d = \kappa_{\text{SLM}} - \tilde{\omega}n$ is the distance of the SLM to the light line, and $K_{s} = \left| \boldsymbol{K}_{(10)} \cdot \boldsymbol{\hat{s}}\right| = \left| \boldsymbol{K}_{(11)} \cdot \boldsymbol{\hat{s}}\right|$ as in sec. \ref{simple model}.

We can then quantify the sensitivity of both the position and the width of the SLR to the inclusion of non-resonant multipoles by one simple parameter $\mathfrak{s} \equiv (\Gamma^{\text{d}}-\Gamma^{\text{cp}})/\Gamma^{\text{d}}$. Here
$\Gamma^{\text{d}}$ is the width parameter predicted if only the dominant multipole is included in the analysis -- in our problem the electric dipole moment for the dipolar SLM and the electric quadrupole moment for the multipolar SLM -- and is found by using $\boldsymbol{\kappa}_{\text{SLM}}$ in that limit in (\ref{Gamma}).
Similarly, $\Gamma^{\text{cp}}$ is the width parameter when the non-resonant multipole -- in our problem the magnetic dipole moment -- is also taken into account, and is found by using the $\boldsymbol{\kappa}_{\text{SLM}}$ of that full calculation in (\ref{Gamma}). Using (\ref{Gamma}) to determine these quantities and noting that the term in brackets varies little when the magnetic dipole is included, we arrive at a simple relation
\begin{equation}
\mathfrak{s} \approx 1 - \frac{d^{\text{cp}}}{d^{\text{d}}},\label{sens}
\end{equation}
where $d^{\text{d}}$ is the distance in reciprocal space of
the position of the SLM to the light line when the magnetic
dipole is neglected, and $d^{\text{cp}}$ is the corresponding distance when the magnetic dipole is included in the calculation. The sensitivity of the SLR to the inclusion of the magnetic dipole moment of the spheres is thus most enhanced when, as a result of the inclusion of the magnetic dipole moment in the calculation, the position of the associated SLM reaches the light line, $d^{\text{cp}}\rightarrow 0$. Thus the merging of the dispersion relation of the multipolar SLM with the light line drives an extreme sensitivity of the multipolar SLR to the inclusion of the magnetic dipole moment. Such an extreme sensitivity is not observed for the dipolar SLR, since the merging of its dispersion relation with the light line is pre-empted by the appearance of the new diffracted order.

The SLM dispersion relations are also indicators of the sensitivity of  SLRs to illumination conditions. This point we discuss in the next section.

\section{Dependence of the spectrum on illumination conditions}
\label{other}

We now turn to the dependence of the reflectance on the polarization and direction of the incident light. We first analyze an angular scan of the reflection with an in-plane wave vector $\boldsymbol{\kappa}_0\propto \boldsymbol{\hat{\kappa}}_a$, as done in sec. \ref{couplings}-\ref{simple model}, but now with the incident light p-polarized. While for s-polarization the two independent sets of moments were $(p_s, m_{\kappa}, q_{\kappa s} )$ and $(m_{\kappa}, q_{sz})$ (see eqs. (\ref{S1},\ref{S2})), for p-polarization they are $(p_{\kappa}, q_{\kappa\kappa}, q_{ss})$ and $(p_z, m_s, q_{\kappa z} )$. That is, both sets involve an electric dipole moment component, coupled to higher multipoles. The structure of the coupling between the multipoles within these sets is important in understanding the radiation from the array. For s-polarized light we could identify qualitatively different SLRs, a dipolar SLR associated with the set of moments $(p_s, m_{\kappa}, q_{\kappa s} )$ and a multipolar SLR associated with the set $(m_{\kappa}, q_{sz})$, but we shall see that the situation is more complicated here. 

\begin{figure}[htb]
\includegraphics[scale=0.3]{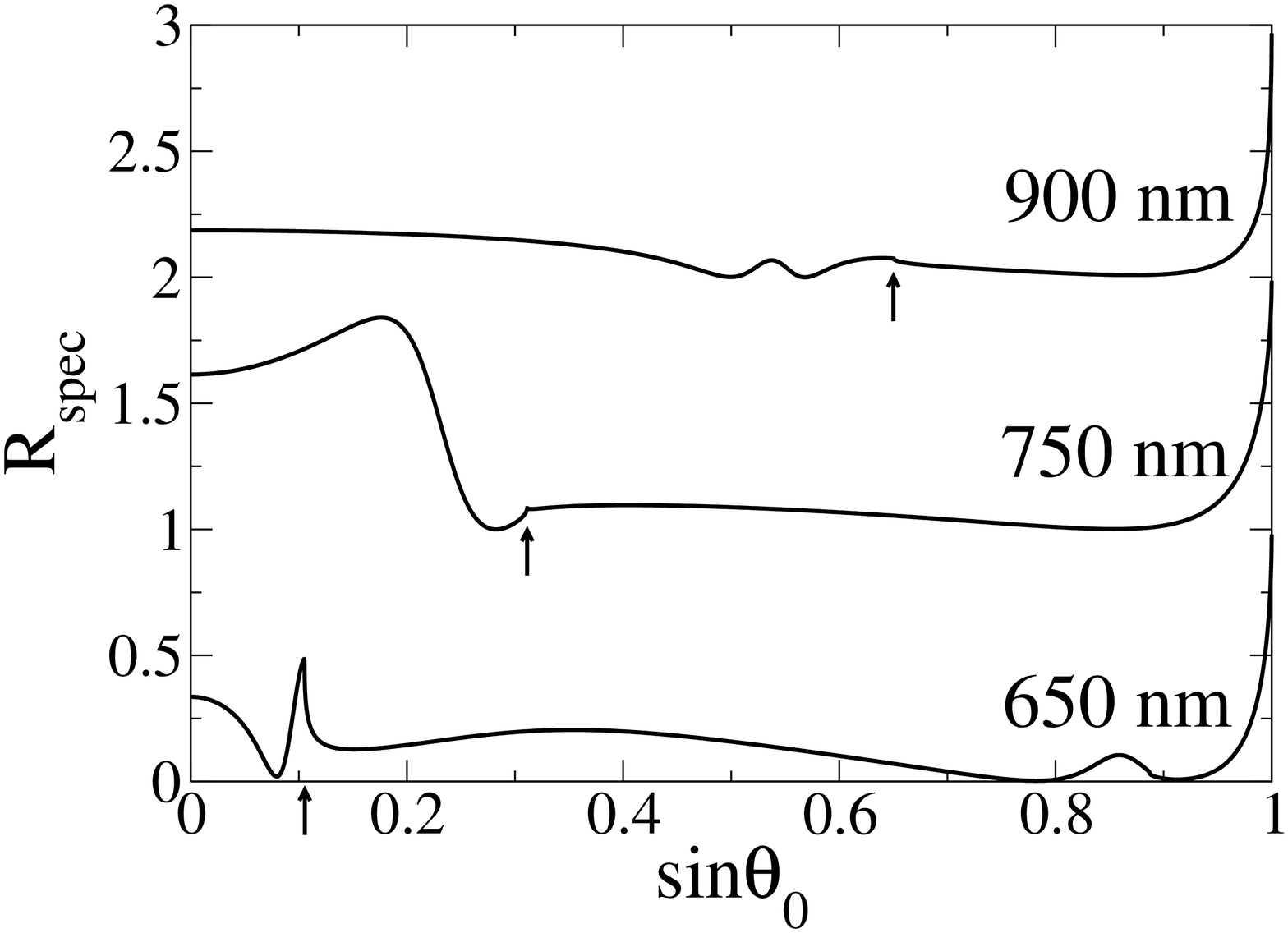}
\caption{Resonant structure of the specular reflection for p-polarized light with $\boldsymbol{\kappa}_0\propto \boldsymbol{\hat{\kappa}}_a$. Arrows indicate the angle at which the Rayleigh line is crossed and diffraction appears. Plots are shifted by 1 to improve clarity.}
\label{ppolarized}
\end{figure}

As in our discussion of illumination by s-polarized light, the Rayleigh line and the light line are identified in Fig.~\ref{ldisp}, and we consider the range of wavelengths between $\lambda_{\text{nr}}$ and $\lambda_{\text{gz}}$. The angular scan of the specular reflectance for wavelengths in this
range is shown in Fig.~\ref{ppolarized}. 
At angles before the onset of diffraction, $\theta_0<\theta_R$, the reflection profile exhibits damped SLRs at long wavelengths, $\lambda \gtrsim 780nm$, a clear broad SLR at intermediate wavelengths, $\lambda \approx (720nm-780nm)$, and some resonant structure or kinks in reflectivity at short wavelengths, $\lambda \lesssim 720nm$. To gain insight into the nature of these resonances we analyse the contributions to reflection spectrum from the two sets of moments. From a calculation within electric dipole approximation we find that at long wavelengths the response of both the in-plane ($p_{\kappa}$) and out-of-plane ($p_z$) components of the electric dipole moment are resonant.  The broad SLRs found within the full multipolar model at long wavelengths are of mostly dipole character and result from a mostly destructive interference of the resonances associated with the sets $(p_{\kappa}, q_{\kappa\kappa}, q_{ss})$ and $(p_z, m_s , q_{\kappa z})$. At intermediate wavelengths the response of the electric dipole component $p_{\kappa}$ is resonant while $p_z$ is not, and the reflection profile is dominated by an SLR associated with the set of moments $(p_{\kappa}, q_{\kappa\kappa}, q_{ ss})$. At short wavelengths the response of neither $p_{\kappa}$ nor $p_z$ is resonant. Here the SLRs are of a mixed electric dipole-electric quadrupole character in a narrow range of wavelengths close to $\lambda\approx 650nm$, and kinks in reflection arise at other short wavelengths. But generally there is no clear SLR structure that can be associated with higher order multipole moments, in contrast to the prominent structure of multipolar SLRs found in the angular scans for s-polarized incident light, as shown in sec. \ref{couplings}.

To investigate this difference, we first compare the effects on the specular reflection of s- and p-polarized light due to the sets of multipoles $(p_s, m_z, q_{\kappa s})$ and $(p_{\kappa}, q_{\kappa \kappa}, q_{ss})$, excited by
s-polarized light and p-polarized light respectively (see Eqs. (\ref{S1}-\ref{P2})); both of these involve an in-plane component of the dipole moment coupled to higher multipoles. The similar nature of the matrices $\mathbb{S}$ and $\mathbb{P}$ that govern these sets of multipoles results in qualitative similar angular scans of specular reflection, dominated by a broad SLR of dipolar nature over a wide range of wavelengths. In contrast, the natures of the matrices $\mathbb{S}'$ and $\mathbb{P}'$ governing of the response of the sets of multipoles $(m_{\kappa}, q_{sz})$ and $(p_z, m_s, q_{\kappa z} )$, excited by s-polarized light and p-polarized light respectively, are quite different; the higher multipoles in the set $(p_z, m_s, q_{\kappa z})$ are coupled to a dipole component, while no such component is present in the set $(m_{\kappa}, q_{sz})$. Not surprisingly, the angular scans of the specular reflection due to the two sets of multipoles are markedly different at the long wavelengths at which the response of $p_z$ is resonant. At these wavelengths the contribution to the specular reflection due to the multipoles $(p_z , m_s , q_{\kappa z})$ leads to a broad response, much like the dipolar SLR that arises due to the moments $(p_s , m_z , q_{\kappa s} )$ that are excited by s-polarized light, while no dipole moment component is present in the set of multipoles $(m_{\kappa}, q_{sz})$. But what is surprising is the markedly different angular scans of specular reflection due to the sets of multipoles $(p_z , m_s , q_{\kappa z} )$ and $(m_{\kappa} , q_{sz})$, excited by p-polarized light and s-polarized light respectively, at intermediate and short wavelengths at which the response of the dipole moment component $p_z$ is non-resonant.

The difference arises due to the coupling of the higher multipoles in the set $(p_z, m_s, q_{\kappa z})$ to the non-resonant electric dipole component, as can be seen by comparing the radiation from this set with that which would be predicted were the dipole component $p_z$ neglected in the calculation. In the latter calculation there is a clear structure with sharp features, similar to the multipolar SLR due to the set of moments $(m_{\kappa}, q_{sz})$ observed
in the response of the array to s-polarized light; this structure is absent in the full calculation of the radiation of the moments $(p_z , m_s, q_{\kappa z})$. This extreme sensitivity of the multipolar response to changes in multipolar couplings can be understood by analyzing the SLMs of the array. We come back to this at the end of this section.

\begin{figure}[htb]
\hspace{-50mm}
\begin{tabular}{p{0.08\textwidth} p{0.08\textwidth}}
\vspace{0pt} \includegraphics[width=0.08\textwidth]{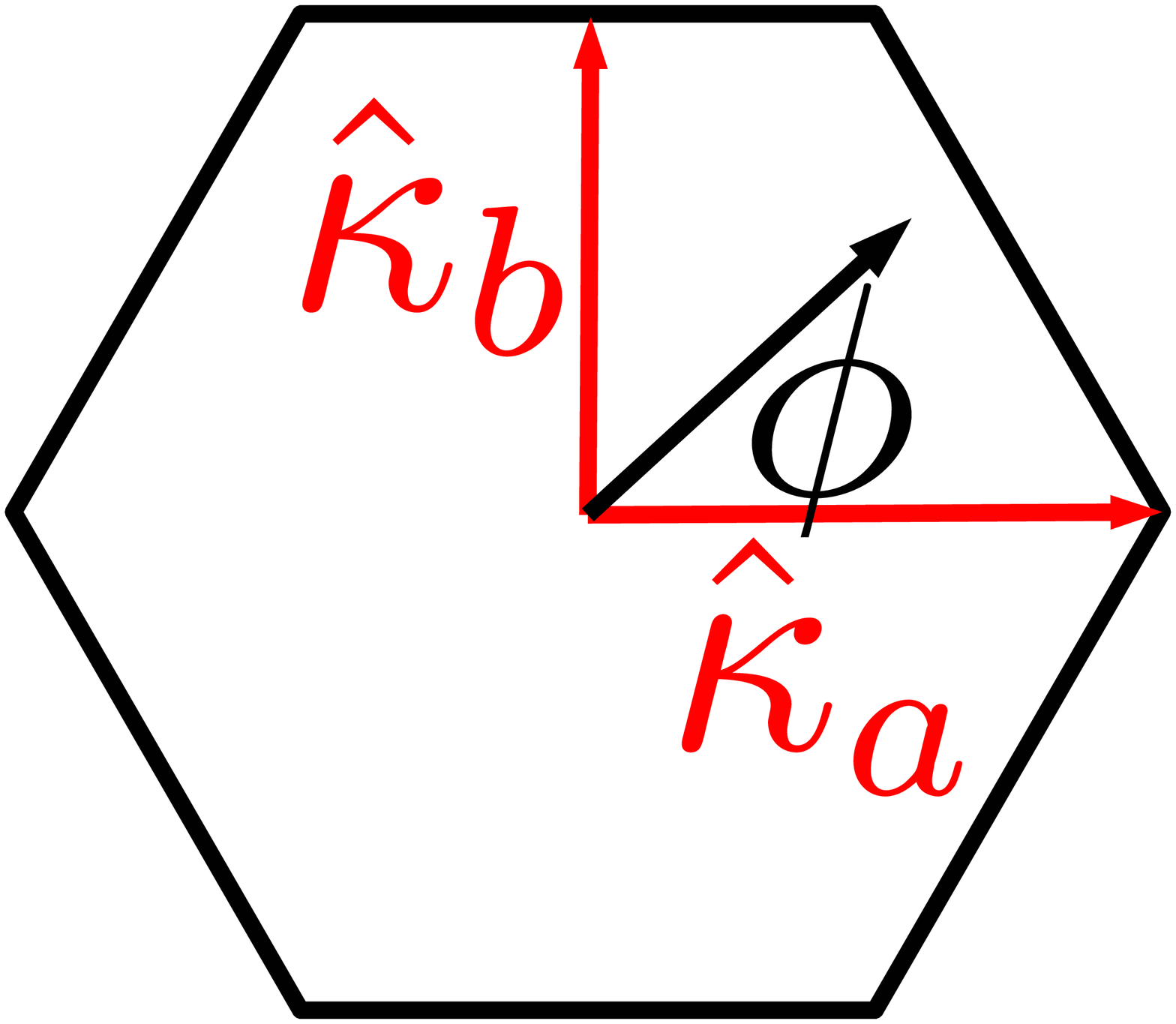} &
\vspace{0pt} \includegraphics[width=0.35\textwidth]{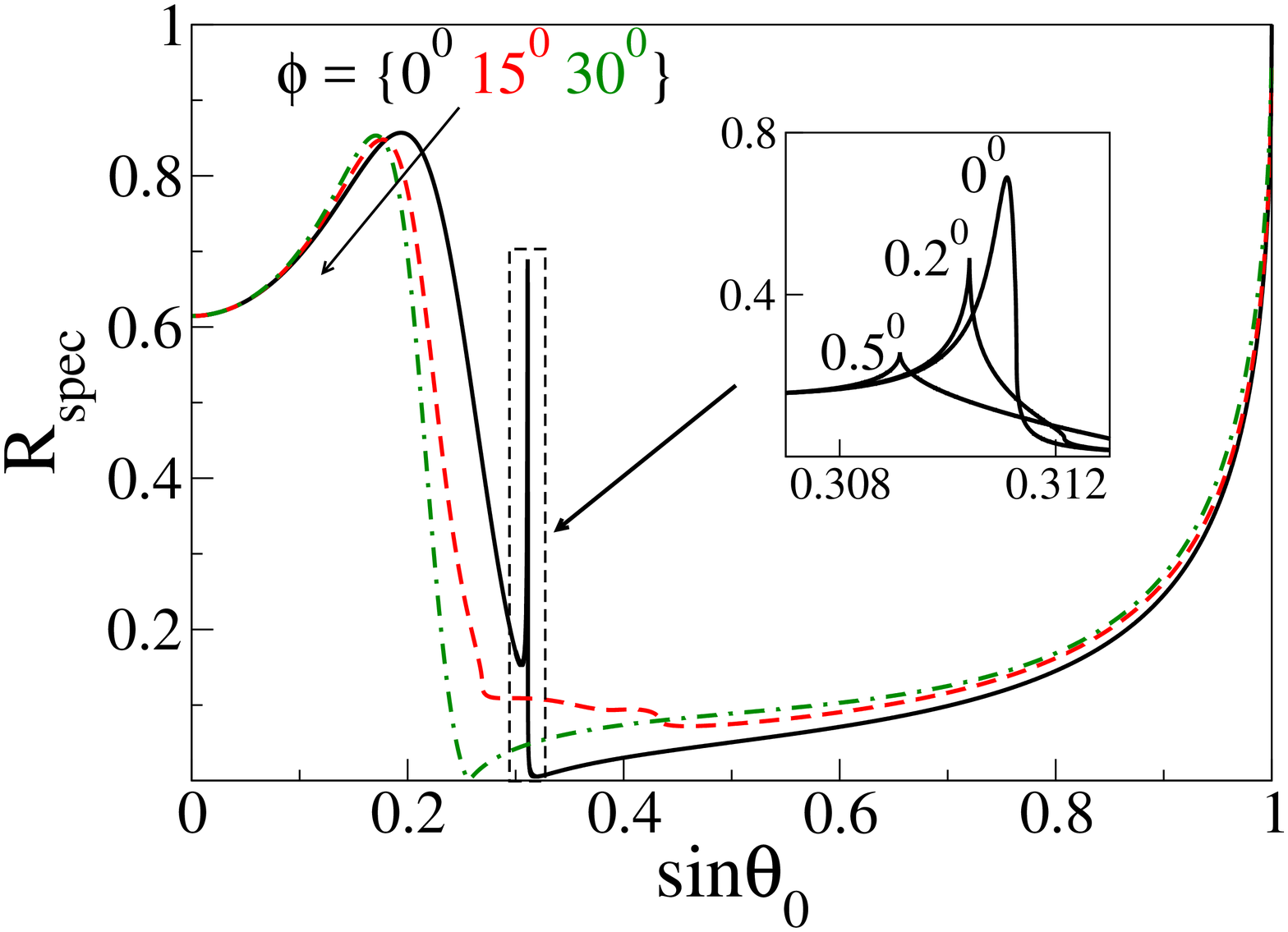}
\end{tabular}
\caption{Specular reflection for s-polarized light at $\lambda =750nm$ for different values of $\phi$. Inset on the left shows the first Brillouin zone with the two symmetry directions indicated by $\boldsymbol{\hat{\kappa}}_a$ and $\boldsymbol{\hat{\kappa}}_b$; compare with the inset in Fig.~\ref{symmetry}, where the second Brillouin zone is also indicated. Inset on the right shows the disappearance of multipolar SLR at small angle $\phi$.}
\label{angles}
\end{figure}

We now turn to a consideration of how the direction of the in-plane wave vector $\boldsymbol{\kappa}_0$ affects the angular scans of specular reflectance. The most significant change of reflectance with the direction of $\boldsymbol{\kappa}_0$ is on the sharp features associated with the multipolar SLR for s-polarized illumination, and since for p-polarized illumination no such sharp features arise we restrict our attention for the rest of this section on s-polarized illumination. We identify the direction of $\boldsymbol{\kappa}_0$ by the angle that it makes with that
symmetry direction, $\phi = \text{cos}^{-1}\left( \boldsymbol{\hat{\kappa}}_0\cdot \boldsymbol{\hat{\kappa}}_a \right)$, and we refer to $\phi$ as a \textit{misalignment} angle; see Fig.~\ref{angles}. As the lattice has a 6-fold rotational symmetry and an inversion symmetry, we need only consider angles in the interval $\phi\in (0^0,30^0)$; here $\phi=0$ corresponds to wave vector along the first symmetry direction, $\boldsymbol{\kappa}_0 \propto \boldsymbol{\hat{\kappa}}_a$, and $\phi=30^0$ is equivalent to wave vector along the second symmetry direction, $\boldsymbol{\kappa}_0 \propto \boldsymbol{\hat{\kappa}}_b$. We calculate angular scans of the specular reflectance by keeping the misalignment angle fixed and varying the angle of incidence.

These angular scans reveal that, while the broad dipolar SLR is only weakly dependent on the direction of $\boldsymbol{\kappa}_0$, the multipolar SLR is strongly anisotropic and observed only when $\boldsymbol{\kappa}_0$ is closely aligned with the symmetry direction of the lattice; see Fig.~\ref{angles} for a comparison at $\lambda = 750nm$. 
To compare the anisotropy of multipolar response at different wavelengths we quantify it by the value of the misalignment angle at which the multipolar SLR merges with the Rayleigh anomaly and vanishes; we denote this angle by $\phi_{\text{an}}$. At long wavelengths the multipolar response is extremely anisotropic, $\phi_{\text{an}}\approx 0.1^0$ at $\lambda\approx 750nm$. At shorter wavelengths the anisotropy is slightly less pronounced, $\phi_{\text{an}}\approx 5^0$ at $\lambda \approx 650nm$. Nevertheless, the multipolar response of the array is strongly anisotropic over the whole range of wavelengths that we consider. This strong anisotropy is a result of sensitivity of the multipolar response to the changes in the structure of couplings between multipole moments. That structure changes qualitatively when the misalignment angle is varied from $\phi=0$ to a small finite value, with the multipole moments decoupled into two independent sets at $\phi=0$, but all the excited multipoles being coupled together for incidence direction misaligned with the lattice symmetry direction. Yet the degree of anisotropy that we observe is surprising, as we might expect the additional couplings that arise at small values of $\phi$  to be weak.

Finally we note that at $\phi = 30^0$ and equivalent angles (including along $\boldsymbol{\hat{\kappa}}_b$) the multipolar SLR is absent; see the reflection spectrum at $\phi = 30^0$ in Fig.~\ref{angles}. At $\phi=0$ and $\phi=30^0$ the multipole moments are described by the same set of equations (\ref{S1}-\ref{S2}) and hence the structure of the couplings is the same. However, for the the wave vector oriented at an angle $\phi=30^0$, the periodic Green functions that are associated with the higher multipole moments $(m_{\kappa}, q_{sz})$ do not diverge at the onset of a new diffraction order, and the multipolar SLRs do not appear. This is associated with the transverse nature of the radiated electromagnetic field,\cite{Mousavi2011} and is easiest seen for the radiation from $m_{\kappa}$. We shall see in the next section that the diffracted field that arises first for
incidence along $\boldsymbol{\hat{\kappa}}_b$ (or an equivalent direction) is characterized by an in-plane wave vector opposite that of the incident field, and at diffraction onset the diffracted amplitude must vanish then since a dipole cannot radiate in the direction in which it points.

We have shown that multipolar response is very sensitive to even small changes in the structure of the couplings between multipoles that are driven by changes in either the polarization or in the direction of incident light. This sensitivity can be understood by analyzing SLMs of the array. Consider first the SLMs excited with s-polarized light incident along $\boldsymbol{\hat{\kappa}}_a$; see Fig.~\ref{disp} in sec. \ref{normal modes}. The  multipolar SLM is in a close proximity to the light line. Thus even a small change in the dispersion relation due to a misalignment of $\boldsymbol{\kappa}_0$ with the symmetry direction $\boldsymbol{\hat{\kappa}}_a$ results in the merging of the SLM with the light line and a disappearance of the SLR associated with it. The dipolar dispersion relation is further away from the light line, and a change in the dipolar dispersion relation results in a shift of the SLM rather than its termination, with the associated SLR of dipolar character existing over a wide range of illumination conditions. A similar reasoning explains the suppression of the multipolar response for p-polarized light when the coupling of the higher multipoles to non-resonant electric dipole is taken into account. Were the coupling to the electric dipole component $p_z$ neglected, a SLM associated with the set $(p_z, m_s, q_{\kappa z} )$ would be found over a wide range of wavelengths, at wave vectors in a close proximity to the light line; when the coupling to $p_z$ is taken into account the dispersion relation merges with the light line and the SLM is not found.

\section{Direct coupling to surface-lattice modes of the array}
\label{substrate}

In the previous sections we have discussed a $\textit{diffractive}$ coupling to SLMs of the array that is signalled by the appearance of SLRs in reflection. Yet the physical significance we have attributed to the SLMs leads immediately to the suggestion that it should be possible to construct a configuration where \textit{direct} coupling to them would be possible. We do that here, and investigate the signatures of direct coupling on an angular scan of the specular reflectance.

\begin{figure}[htb]
\includegraphics[scale=0.25]{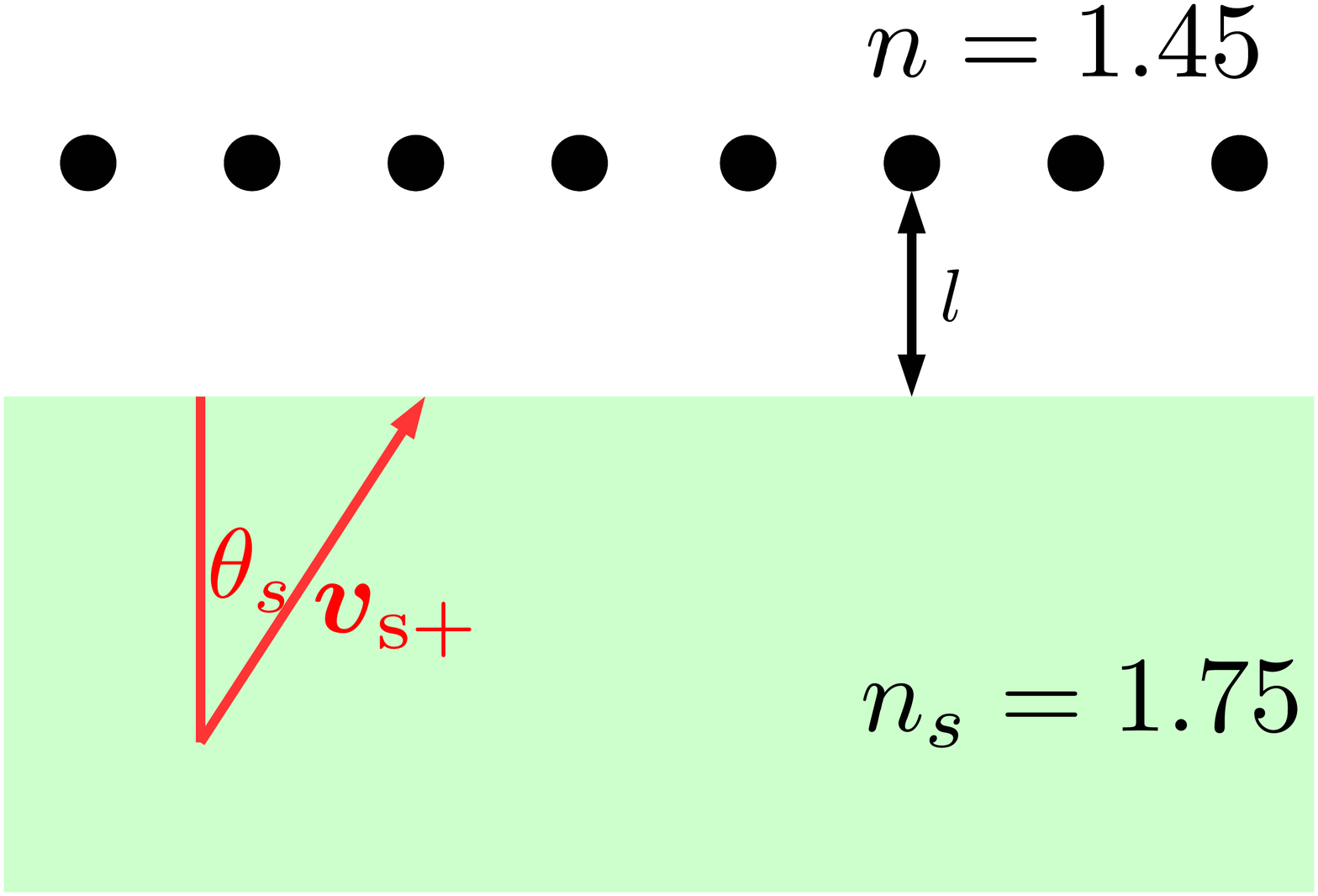}
\caption{The array at a distance $l$ above an interface. The interface is  between two semi-infinite homogenous media, with the background having the index of refraction $n=1.45$, and the substrate having the index of refraction $n_s=1.75$. The array is illuminated by a
plane wave incident from the substrate.}
\label{sphsub}
\end{figure}

Again the analogy with a thin metal film is a useful one. Suppose we have a metal film embedded in a background medium with index of refraction $n$ above a substrate with a higher index of refraction, $n_s>n$; see \mbox{Fig.~\ref{sphsub},}
but with the array of spheres imagined replaced by a metal film. When the light is incident from the substrate above the critical angle,  $\tilde{\omega}n_s>\kappa_{\text{inc}}>\tilde{\omega}n$, where $\boldsymbol{\kappa}_{\text{inc}}$ is the in-plane component of the  wave vector of light, a long-range surface plasmon\cite{Berini} (LRSP) could be excited by electromagnetic field that is evanescent in the background medium. Such an excitation would lead to absorption loss, and so a dip in the reflectivity would be seen at the angle where the coupling to the LRSP is optimal. Of course, the  $\boldsymbol{\kappa}_{\text{inc}}$ where the dip would occur would not be determined exactly by the dispersion relation of the LRSP in the absence of the substrate -- the presence of the substrate would modify the dispersion relation -- but it would be close if the distance between the substrate and the  metal film were large enough that the coupling were weak.

With this analogy in mind, we consider a 2D array of spheres a distance $l$ above an interface at $z = 0$ separating two semi-infinite homogeneous media, where the index of refraction for the background medium at $z > 0$ is $n = 1.45$, and for the substrate at $z < 0$ it is $n_s = 1.75$. Suppose that when free-standing in the background medium the array exhibits an SLR at $\boldsymbol{\kappa}_{\text{SLR}}$, which signals coupling into a SLM at $\boldsymbol{\kappa}_{\text{SLM}} = \boldsymbol{\kappa}_{\text{SLR}} + \boldsymbol{K}$, where $\boldsymbol{K}$ is one of the reciprocal lattice vectors, and of course $|\boldsymbol{\kappa}_{\text{SLR}}| < \tilde{\omega}n$. In the presence of the substrate the SLM is shifted from its position $\boldsymbol{\kappa}_{\text{SLM}}$ for a free-standing array, but if that shift is small a field incident from the substrate with an in-plane wave vector $\boldsymbol{\kappa}_{\text{inc}}$ could certainly couple diffractively into the SLM when $\boldsymbol{\kappa}_{\text{inc}}\approx \boldsymbol{\kappa}_{\text{SLR}}$. But in addition to the diffractive coupling, a direct coupling to the SLM should also be possible at an in-plane wave vector $\boldsymbol{\kappa}_{\text{inc}}\approx \boldsymbol{\kappa}_{\text{SLM}}$ that corresponds to incidence at an angle above the critical angle, $|\boldsymbol{\kappa}_{\text{inc}}| > \tilde{\omega}n$.

For simplicity and to illustrate the ideas more clearly we consider s-polarized light incident from the substrate with a wave vector component $\boldsymbol{\kappa}_{\text{inc}}$ along the second symmetry direction $\hat{\boldsymbol{\kappa}}_{b}$ (see Figure \ref{symmetry}). We consider a plane wave incident from the substrate,
\begin{align}
 & \boldsymbol{E}^{\text{inc}} (\boldsymbol{r})=\hat{\boldsymbol{s}}E^{\text{inc}}  e^{i\boldsymbol{\nu}_{s+}\cdot\boldsymbol{r}},
\end{align}
where $\boldsymbol{\kappa}_{\text{inc}}=\tilde{\omega}n_s \boldsymbol{\hat{\kappa}}_b\text{sin}\theta_s$, $\boldsymbol{\hat{s}} = \boldsymbol{\hat{\kappa}}_{\text{inc}}\times\boldsymbol{\hat{z}}$, and $\boldsymbol{\nu}_{s\pm}=\boldsymbol{\kappa}_{\text{inc}}\pm w_{s}(\kappa_{\text{inc}})\hat{\boldsymbol{z}}$ with $w_{s}(\kappa_{\text{inc}})=\sqrt{\tilde{\omega}^2n_s^{2} -\kappa_{\text{inc}}^{2}}$. In the absence of the array of spheres, the fields that would be reflected and transmitted are given by
\begin{align}
\boldsymbol{E}^R(\boldsymbol{r}) &= r^s_0 E^{\text{inc}} \boldsymbol{\hat{s}} e^{i\boldsymbol{v}_{s-}\cdot \boldsymbol{r}}, \label{reflected}\\
\boldsymbol{E}^T(\boldsymbol{r}) &= t^s_0 E^{\text{inc}} \boldsymbol{\hat{s}} e^{i\boldsymbol{v}_{+}\cdot \boldsymbol{r}},\label{transmitted}
\end{align}
where $r^s_0$ and $t^s_0$ are the usual Fresnel coefficients for s-polarized light, 
\begin{align}
r^s_0 &= \frac{w_s(\kappa_{\text{inc}}) - w_0(\kappa_{\text{inc}})}{w_s(\kappa_{\text{inc}})+w_0(\kappa_{\text{inc}})}, \label{rs0}\\
t^s_0 &= \frac{2w_s(\kappa_{\text{inc}})}{w_s(\kappa_{inc})+w_0(\kappa_{\text{inc}})},
\end{align}
with $w_s(\kappa_{\text{inc}})$ as above, and $w_0(\kappa_{\text{inc}}) = \sqrt{\tilde{\omega}^2 n^2 - \kappa^2_{\text{inc}}}$. 

We now include the presence of the array of spheres shown in Fig.~\ref{sphsub}. For our excitation scenario the optical response of the spheres is well described within the electric dipole approximation over a wide range of wavelengths, since as discussed in sec. \ref{other} the narrow SLR associated with higher order multipole moments is not found. We thus model the spheres as electric dipoles. Using the phase relation between the dipole moments, eq. (\ref{rel1}), we formulate the response of the array in terms of the dipole $\boldsymbol{p}$ at the origin of the lattice, $\boldsymbol{R}=0, z=l$. The dipole is driven by an electric field that can be written as
the sum of three terms: (a) the transmitted field  (\ref{transmitted}) at the position of
the dipole that would arise were the array not present, $\boldsymbol{E}^T = \boldsymbol{E}^T(\boldsymbol{R}=0,l)$, (b) a sum of the radiation reaction field and the field scattered from all the other spheres, $\ovb{\mathcal{G}}^{\text{Ep}}\cdot\boldsymbol{p}$, where $\ovb{\mathcal{G}}^{\text{Ep}}$ is the periodic Green function introduced earlier, (c) the field that is radiated by the array and reflected back from the substrate, $\ovb{\mathcal{G}}_s^{\text{Ep}} \cdot\boldsymbol{p}$, where $\ovb{\mathcal{G}}_s^{\text{Ep}}$ is the substrate correction to the dipole Green function,\cite{Sipe1987}
\begin{align}
\ovb{\mathcal{G}}_s^{\text{Ep}} = \frac{1}{A_c} \sum_{{\boldsymbol{n}}} \frac{i\tilde{\omega}^2}{2\epsilon_0} \frac{ e^{2i w_0(|\boldsymbol{\kappa}_{\text{inc}} + \boldsymbol{K}_{\boldsymbol{n}}|) l}}{ w_0(|\boldsymbol{\kappa}_{\text{inc}} + \boldsymbol{K}_{\boldsymbol{n}}|)} \mathcal{A}_{\boldsymbol{n}}, \label{subG}
\end{align}
where we have defined a dyadic 
\begin{equation}
\mathcal{A}_{\boldsymbol{n}} = r^s_{\boldsymbol{n}} \boldsymbol{\hat{s}}_{\boldsymbol{n}} \boldsymbol{\hat{s}}_{\boldsymbol{n}} + r^p_{\boldsymbol{n}} \boldsymbol{\hat{p}}_{\boldsymbol{n}+}\boldsymbol{\hat{p}} _{\boldsymbol{n}-},
\end{equation}
with 
\begin{align}
\boldsymbol{\hat{s}}_{\boldsymbol{n}} &= |\boldsymbol{\kappa}_{\text{inc}} + \boldsymbol{K}_{\boldsymbol{n}}|^{-1} (\boldsymbol{\kappa}_{\text{inc}} + \boldsymbol{K}_{\boldsymbol{n}} ) \times \boldsymbol{\hat{z}}, \\
\boldsymbol{\hat{p}}_{\boldsymbol{n}\pm} &= (\tilde{\omega}n)^{-1}|\boldsymbol{\kappa}_{\text{inc}} + \boldsymbol{K}_{\boldsymbol{n}}|\boldsymbol{\hat{z}} \nonumber\\
&\mp (\tilde{\omega}n)^{-1} w_0(|\boldsymbol{\kappa}_{\text{inc}} +  \boldsymbol{K}_{\boldsymbol{n}}|) \frac{(\boldsymbol{\kappa}_{\text{inc}} + \boldsymbol{K}_{\boldsymbol{n}})\times \boldsymbol{\hat{z}}}{|\boldsymbol{\kappa}_{\text{inc}} + \boldsymbol{K}_{\boldsymbol{n}}|^{-1}}.  
\end{align}
Here $r^s_{\boldsymbol{n}}$ is the s-polarized reflection coefficient (see eq. (\ref{rs0})) with $\kappa_{\text{inc}}$ replaced by $\left|\boldsymbol{\kappa}_{\text{inc}}+\boldsymbol{K}_{\boldsymbol{n}} \right|$, and  $r^p_{\boldsymbol{n}}$ is the corresponding p-polarized  Fresnel reflection coefficient,
\begin{equation}
r^p_{\boldsymbol{n}} = \frac{n^2 w_s\left(|\boldsymbol{\kappa}_{\text{inc}}+\boldsymbol{K}_{\boldsymbol{n}}|\right) - n_s^2w_0\left(|\boldsymbol{\kappa}_{\text{inc}}+\boldsymbol{K}_{\boldsymbol{n}}|\right)}{n^2w_s\left(|\boldsymbol{\kappa}_{\text{inc}}+\boldsymbol{K}_{\boldsymbol{n}}|\right)+n_s^2w_0\left(|\boldsymbol{\kappa}_{\text{inc}}+\boldsymbol{K}_{\boldsymbol{n}}|\right)}.
\end{equation}
Adding all the contributions we arrive at the total driving field,
\begin{align}
\boldsymbol{E}^{\text{tot}} &= \boldsymbol{E}^T + \left[ \ovb{\mathcal{G}}^{\text{Ep}} + \ovb{\mathcal{G}}_s^{\text{Ep}} \right]\cdot \boldsymbol{p}\label{EtotS},
\end{align}
in terms of which the electric dipole is given by
\begin{equation}
\boldsymbol{p} = \alpha^{\text{pE}} \boldsymbol{E}^{\text{tot}}.\label{subp}
\end{equation}

\begin{figure}[htb]
\includegraphics[scale=0.25]{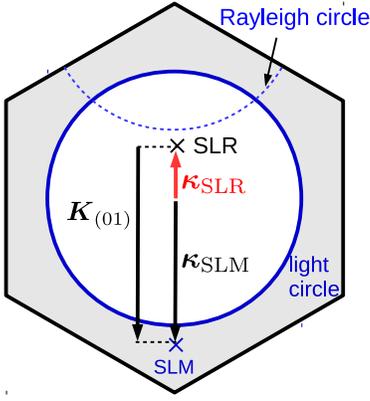}
\caption{First and second Brillouin zone of the reciprocal lattice. For illumination with in-plane wave vector $\boldsymbol{\kappa}_{\text{inc}}$ along $\boldsymbol{\hat{\kappa}}_b$, the SLR is found at an in-plane wave vector $\boldsymbol{\kappa}_{\text{SLR}}$ and the SLM at  $\boldsymbol{\kappa}_{\text{SLM}} = \boldsymbol{\kappa}_{\text{SLR}}+\boldsymbol{K}_{(01)}$.}
\label{invsub}
\end{figure}

We now calculate an angular scan of the specular reflectance of the structure at $\lambda=900nm$. For $\boldsymbol{\kappa}_{\text{inc}}$ in the direction of $\boldsymbol{\hat{\kappa}}_b$ the first reciprocal lattice vector of importance as $|\boldsymbol{\kappa}_{\text{inc}}|$ increases from zero is  $\boldsymbol{K}_{(01)}$; see Fig.~\ref{invsub}.
We choose the  array to be at a distance $l=l_{0}$ from the interface, where $l_{0}=\left(\kappa^{2}_{\text{SLM}} -\tilde{\omega}^{2}n^{2}\right)^{-1/2} = 337.2nm$
with $\boldsymbol{\kappa}_{\text{SLM}} = \boldsymbol{\kappa}_{\text{SLR}}+\boldsymbol{K}_{(01)}$ being the wave vector of the SLM supported by the free-standing array. For this choice of the distance, the evanescent fields above the substrate at angles of incidence beyond the critical angle excite clear resonances in the array, yet the array is far enough from the interface so that its mode structure is not significantly modified.

\begin{figure}[htb]
\includegraphics[scale=0.3]{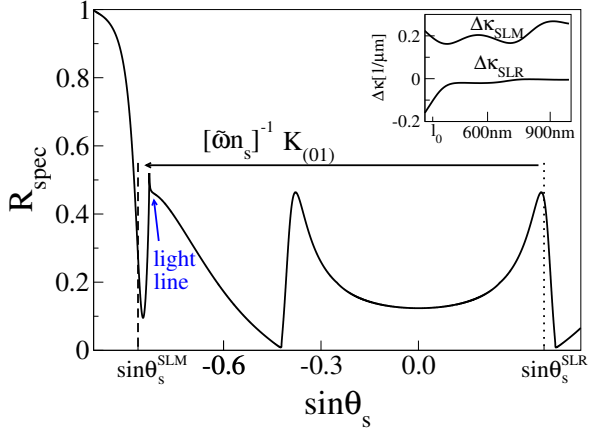}
\caption{Specular reflection from the array at a distance $l_{0}= 337.2nm$ from an interface, at $\lambda=900nm$. The dashed and dotted lines indicate the angles at which incident light would couple directly and with the help of the grating to the SLM of the array if it were free-standing in a background medium. The inset shows deviations from these angles of the resonance features, measured as a distance in reciprocal space, as $l$ increases from its value at $l_{0}$. }
\label{substrate1}
\end{figure}

The specularly reflected light exhibits two resonant features of a distinct character as the angle of incidence in the substrate, $\theta_s$, departs from zero. As $\theta_s$ becomes increasingly positive, we see an SLR-like peak at an angle near that which corresponds to the SLR resonance of the
free-standing array, $\text{sin}\theta_s^{\text{SLR}} = \kappa_{\text{SLR}}/\tilde{\omega}n_s$; we indicate this angle as a dotted line in Fig.~\ref{substrate1},
and it corresponds to diffractive coupling to the SLM, at an  angle close to what is seen in the specular reflection of a free-standing array. From Fig.~\ref{invsub} we expect the
SLM associated with this SLR to be at $\boldsymbol{\kappa}_{\text{SLM}} = \boldsymbol{\kappa}_{\text{SLR}} + \boldsymbol{K}_{(01)}$, which is accessible to light incident from the prism at a negative angle $\theta_s^{\text{SLM}}$, $\text{sin}\theta_s^{\text{SLM}} = - \kappa_{\text{SLM}}/\tilde{\omega}n_s$. We indicate this angle as a dashed line in Fig.~\ref{substrate1}, and near this there is a dip in the specular reflectance analogous to the dip one sees in surface plasmon studies, for example, where the excitation of the surface mode leads to a large absorption and a decrease in the reflectance; this corresponds to direct coupling
to the SLM. The third peak that appears in the angular range
shown in Fig.~\ref{substrate1} is near $-\theta_s^{\text{SLR}}$, and corresponds to the SLR that exists from symmetry at $−\kappa_{\text{SLR}}$; associated with it is a dip at $\theta_s^{\text{SLM}}$, not in the angular range shown, which corresponds to direct coupling to the SLM at $-\boldsymbol{\kappa}_{\text{SLM}}$.

The agreement between the resonant structures at angles near $\theta_s^{\text{SLR}}$ and $\theta_s^{\text{SLM}}$ and those angles themselves is not exact. This is not surprising, as the interaction between the array and the substrate modifies the effective polarizability of the array, 
\begin{equation}
\left(\ovb{\alpha}^{\text{eff}}\right)^{-1} = \left(\ovb{\alpha}^{\text{pE}}\right)^{-1} - \ovb{\mathcal{G}}^{\text{Ep}} - \ovb{\mathcal{G}}^{\text{Ep}}_s,
\end{equation}
cf. eqs. (\ref{EtotS},\ref{subp}), and thus it modifies the dispersion relation of the SLM as well. But as $l$ increases the contributions to $\ovb{\mathcal{G}}_s$ due to in-plane wave vectors associated with evanescent fields in the medium becomes negligible, and we can expect better agreement of the positions of the resonant structures with $\theta_s^{\text{SLR}}$ and $\theta_s^{\text{SLM}}$. In the inset of Fig.~\ref{substrate1} we show the deviation of the in-plane wave vectors associated
with the resonant structures from the respective wave vectors $\boldsymbol{\kappa}_{\text{SLR}}$ and $\boldsymbol{\kappa}_{\text{SLM}}$, which we identify as $\Delta \kappa_{\text{SLR}}$ and $\Delta \kappa_{\text{SLM}}$ respectively. As $l$ increases we see that $\Delta \kappa_{\text{SLR}}$ indeed approaches zero, while $\Delta \kappa_{\text{SLM}}$ oscillates around a finite value. The oscillations are associated with the coupling of the evanescent fields exciting the array to fields propagating away from the array by diffraction; these fields can reflect off the interface, propagate back to the layer, and interact with it again, leading to interference effects. The finite value is a shift in the
position of the dip arising because the angle at which the minimum dip
occurs depends not only on the position of the SLM, but also on
how it is driven. For a given field incident from the substrate, the evanescent field at array will be larger for smaller $|\theta_s|$, closer to the cut-off for total internal reflection. This results in a shift of the position of the dip to smaller $|\theta_s|$. Nonetheless, it is clear that the dip can be associated with the direct excitation of the SLM.

\section{Conclusions}
\label{conclusions}
In conclusion, we have analyzed the multipolar structure of the reflection of radiation from an array of gold spheres of moderate size, and its dependence on frequency and illumination conditions. We included the electric dipole, magnetic dipole, and electric quadrupole moments of the spheres; full numerical calculations confirm that these are sufficient to capture the electrodynamics for the range of parameters we have considered. We have shown the importance of including the response of moments that would be non-resonant and negligible in magnitude were only an isolated sphere considered; such moments are sometimes neglected in treatments of arrays, and at first thought
this might seem reasonable. However, the reflection of an array is dominated by the so-called “surface lattice resonances” (SLRs), which appear at angles close to those where new diffraction orders arise, and it is at these angles that the electromagnetic coupling between different moments in spheres in an array can diverge. Due to this coupling, moments that would be unimportant in the response of an isolated sphere -- in our example the magnetic dipole moment -- can have important consequences on the angular and frequency dependence of the reflectivity of an array of spheres. 

First, the coupling between the moments can suppress the SLRs associated primarily with moments higher than the electric dipole. In the lattice we considered, such an SLR only survives for s-polarized light incident in one symmetry direction, where it is due to a dominant quadrupole moment coupled only to the magnetic dipole moment. The introduction of additional couplings between the moments, due to a change in the polarization or direction of the incident field, results in a vanishing of this SLR.
Second, the profile, position, and the cut-off frequency at which the SLRs vanish is affected by the coupling to non-resonant moments. This is most significant for the SLRs that do not involve the electric dipole moment in a significant way, for due to the weak coupling with the incident light these SLRs are close to the angles at which new diffraction orders arise. But even for the other SLRs the coupling effects are non-negligible. 

When the interaction between the different multipoles are important the analysis of SLRs becomes complicated, but it can be simplified by identifying the ``ideal" surface-lattice modes (SLMs) of the array. They should be contrasted with the waveguide-like excitations that exist with the neglect of absorption at wavelengths too large for diffraction to occur, and with the bound states in the continuum that do not radiate because of symmetry or interference between radiative channels.\cite{BSC} Instead, the ideal surface-lattice modes we consider arise at wavelengths where diffraction occurs, and are identified by neglecting both absorption losses and diffractive losses. The dispersion relations of these ideal SLMs gives an extremely good prediction of where the SLRs should appear, even when absorption and diffraction are included; the physical picture is that diffraction allows coupling into them from the incident beam, and the coupling back leads to the SLR observed in reflection. We have shown that the dispersion relations can be understood within simplified models based on isolating the divergent and leading analytic contributions to the coupling between the moments, and that an understanding of the properties of the associated SLRs then follows. In particular, we identified differences in the effects that drive the termination of different ideal SLM dispersion relations, and have linked them to the difference in the sensitivity of the different SLRs to the couplings between the multipole moments and the illumination conditions. 

Finally, we have shown that it should be possible to observe these ideal SLMs not only by coupling into them through diffraction, where they lead to the SLRs, but as well by directly coupling into them through a high-index prism. This is analogous to the coupling into surface plasmons or long-range surface plasmons in Kretschmann or Otto configurations, and should be a new way to study these SLMs. 

While we have considered only one lattice and one size of spheres, we have looked at the reflectivity as both a function of angle and wavelength over the range of parameters where the onset of diffraction is important, and where the isolated spheres would exhibit some resonant response. We have examined in detail the response for light incident along two high symmetry
directions, one in which the electrodynamics is essentially one-dimensional in its important features, and one in which it is fully two-dimensional, and as well looked at how the situation changes when one moves away from these high symmetry directions. Thus the methods and concepts that we have applied here should be useful more generally to other systems as well.

\begin{acknowledgments}
We acknowledge financial support from the Natural Sciences and Engineering Research Council of Canada.
\end{acknowledgments}

\appendix

\section{}
The expression for the matrices entering the self-consistent equations for the multipole moments for s-polarized light (\ref{S1},\ref{S2}) are given by
\begin{align}
\mathbb{S} &= \mathbb{A}^S  - \mathbb{G}^S, \\
\mathbb{S}'&= \mathbb{A}'^S - \mathbb{G}'^S,
\end{align}
where the matrices involving the proper polarizability tensors are given by
\begin{align*} 
\mathbb{A}^S &= 
\left( 
\begin{array}{ccc}
(\alpha^{pE})^{-1}  & 0 & 0 \\
0 & (\alpha^{mB})^{-1} & 0 \\
0 & 0 & \left(\alpha^{qF} \right)^{-1} 
\end{array} \right), \\
\mathbb{A}'^S &= 
\left( 
\begin{array}{cc}
(\alpha^{mB})^{-1} & 0 \\
0 & \left(\alpha^{qF} \right)^{-1} 
\end{array} \right),
\end{align*}
and the matrices involving the Green functions are
\begin{align*} 
\mathbb{G}^S &= \left( 
\arraycolsep=2.0pt\def\arraystretch{1.5}
\begin{array}{ccc}
 \mathcal{G}^{\text{EP}}_{ss} & \mathcal{G}^{\text{EM}}_{sz} & 2 \mathcal{G}^{\text{EQ}}_{ss\kappa} \\
\mathcal{G}^{\text{BP}}_{zs} & \mathcal{G}^{\text{BM}}_{zz} & 2\mathcal{G}^{\text{BQ}}_{z\kappa s} \\
\mathcal{G}^{\text{FP}}_{\kappa ss} & \mathcal{G}^{\text{FM}}_{\kappa s z} & 2 \mathcal{G}^{\text{FQ}}_{\kappa s\kappa s}
\end{array} \right), \\
\mathbb{G}'^S &= \left( 
\arraycolsep=2.0pt\def\arraystretch{1.5}
\begin{array}{ccc}
 \mathcal{G}^{\text{BM}}_{\kappa \kappa} & 2\mathcal{G}^{\text{BQ}}_{\kappa sz}  \\
\mathcal{G}^{\text{FM}}_{sz\kappa} & 2\mathcal{G}^{\text{FQ}}_{szsz} 
\end{array} \right),
\end{align*}
where the elements take on different values depending on $\kappa$, $\omega$ and on whether $\boldsymbol{\hat{\kappa}}_0 = \boldsymbol{\hat{\kappa}}_a$ or $\boldsymbol{\hat{\kappa}}_0 = \boldsymbol{\hat{\kappa}}_b$. Similarly, for p-polarized light we have 
\begin{align}
\mathbb{P} &= \mathbb{A}^P  - \mathbb{G}^P, \\
\mathbb{P}' &= \mathbb{A}'^P   - \mathbb{G}'^P,
\end{align}
where 
\begin{align*}
\mathbb{A}^P &= 
\left( 
\begin{array}{ccc}
(\alpha^{pE})^{-1} & 0 & 0 \\
0 & (\alpha^{qF})^{-1} & 0 \\
0 & 0 & \left(\alpha^{qF} \right)^{-1} 
\end{array} \right), \\
\mathbb{A}'^P &= \mathbb{A}^S,
\end{align*}
and the matrices involving the Green functions are given by
\begin{align*} 
\mathbb{G}^P &= \left( 
\arraycolsep=2.0pt\def\arraystretch{1.5}
\begin{array}{ccc}
 \mathcal{G}^{\text{EP}}_{\kappa\kappa} & \mathcal{G}^{\text{EQ}}_{\kappa\kappa\kappa}-\mathcal{G}^{\text{EQ}}_{\kappa zz} & \mathcal{G}^{\text{EQ}}_{\kappa ss}-\mathcal{G}^{\text{EQ}}_{\kappa zz} \\
\mathcal{G}^{\text{FP}}_{\kappa\kappa\kappa} & \mathcal{G}^{\text{FQ}}_{\kappa\kappa\kappa\kappa} - \mathcal{G}^{\text{FQ}}_{\kappa\kappa zz} & \mathcal{G}^{\text{FQ}}_{\kappa\kappa ss}-\mathcal{G}^{\text{FQ}}_{\kappa\kappa zz} \\
\mathcal{G}^{\text{FP}}_{ss \kappa} & \mathcal{G}^{\text{FQ}}_{ss\kappa\kappa}-\mathcal{G}^{\text{FQ}}_{sszz} &  \mathcal{G}^{\text{FQ}}_{ssss}-\mathcal{G}^{\text{FQ}}_{sszz}
\end{array} \right), \\
\mathbb{G}'^P &= \left( 
\arraycolsep=2.0pt\def\arraystretch{1.5}
\begin{array}{ccc}
 \mathcal{G}^{\text{EP}}_{zz} & \mathcal{G}^{\text{EM}}_{zs} & 2 \mathcal{G}^{\text{EQ}}_{z\kappa z} \\
\mathcal{G}^{\text{BP}}_{sz} & \mathcal{G}^{\text{BM}}_{ss} & 2\mathcal{G}^{\text{BQ}}_{s\kappa z} \\
\mathcal{G}^{\text{FP}}_{\kappa zz} & \mathcal{G}^{\text{FM}}_{\kappa z s} & 2 \mathcal{G}^{\text{FQ}}_{\kappa z\kappa z}
\end{array} \right),
\end{align*}
and again the elements take on different values depending on $\kappa$, $\omega$ and on whether $\boldsymbol{\hat{\kappa}}_0 = \boldsymbol{\hat{\kappa}}_a$ or $\boldsymbol{\hat{\kappa}}_0 = \boldsymbol{\hat{\kappa}}_b$.

\newpage
\bibliography{References}

\end{document}